\documentclass[preprint2]{aastex}

\usepackage{multicol}
\usepackage{subfigure}
\usepackage{graphicx}
\usepackage{txfonts}
\usepackage{natbib}
\bibpunct{(}{)}{;}{a}{}{,}

\shorttitle{OTELO survey: emission-line flux determination with OSIRIS}
\shortauthors{Lara-L\'opez et al.}

\begin{document}


\title{OTELO survey: optimal emission-line flux determination with OSIRIS/GTC}


\author{M.A. Lara-L\'opez\altaffilmark{1,2}, J. Cepa\altaffilmark{1,2}, H. Casta\~neda\altaffilmark{3,1}, A.M. P\'erez Garc\'{\i}a\altaffilmark{1,2}, A. Bongiovanni\altaffilmark{1,2},  A. Ederoclite\altaffilmark{1,2}, M. Fern\'andez Lorenzo\altaffilmark{1,2}, M. Povi\'c\altaffilmark{1,2}, M. S\'anchez-Portal\altaffilmark{4}, E. Alfaro\altaffilmark{5}, J. Gallego\altaffilmark{6}, J. J. Gonz\'alez\altaffilmark{7}, J. I. Gonz\'alez-Serrano\altaffilmark{8}}

\altaffiltext{1}{Instituto de Astrof\'isica de Canarias, 38200 La Laguna, Tenerife, Spain: mall@iac.es}
\altaffiltext{2}{Departamento de Astrof\'{\i}sica, Universidad de la Laguna, Spain}
\altaffiltext{3}{Departamento de F\'{\i}sica, Escuela Superior de F\'{\i}sica y Matem\'atica, IPN, M\'exico D.F., M\'exico}
\altaffiltext{4}{Herschel Science Center, INSA/ESAC, Madrid, Spain}
\altaffiltext{5}{Instituto de Astrof\'isica de Andaluc\'ia-CSIC, Granada, Spain}
\altaffiltext{6}{Departamento de Astrof\'isica y CC. de la Atm\'osfera, Universidad Complutense de Madrid, Madrid, Spain}
\altaffiltext{7}{Instituto de Astronom\'ia UNAM, M\'exico D.F, M\'exico}
\altaffiltext{8}{Instituto de F\'isica de Cantabria, CSIC-Universidad de Cantabria, Santander, Spain}


\begin{abstract}
Emission-line galaxies are important targets for understanding the chemical evolution of galaxies in the universe. Deep, narrow-band imaging surveys allow to detect and study the flux and the equivalent widths (EW) of the emission line studied. The present work has been developed within the context of the OTELO project, an emission line survey using the Tunable Filters (TF) of OSIRIS, the first generation instrument on the GTC 10.4m telescope located in La Palma, Spain, that will observe through selected atmospheric windows relatively free of sky emission lines. With a total survey area of 0.1 square degrees distributed in different fields, reaching a 5 $\sigma$ depth of $10^{-18}$ erg/cm$^2$/s and detecting objects of EW $<$ 0.3 {\AA}, OTELO will be the deepest emission line survey to date. As part of the OTELO preparatory activities, the objective of this study is to determine the best combination of sampling and  full width at half maximum (FWHM) for the OSIRIS tunable filters for deblending  H$\alpha$ from [{N\,\textsc{ii}}] lines by analyzing the flux errors obtained. We simulated the OTELO data by convolving a complete set of synthetic {H\,\textsc{ii}} galaxies in EW with different widths of the OSIRIS TFs. We estimated relative flux errors of the recovered  H$\alpha$ and [{N\,\textsc{ii}}]$\lambda$6583 lines. We found that, for the red TF, a FWHM of 12 {\AA} and a sampling of 5 {\AA} is an optimal combination that allow deblending  H$\alpha$ from the [{N\,\textsc{ii}}]$\lambda$6583 line with a flux error lower than 20$\%$. This combination will allow estimating SFRs and metallicities using the H$\alpha$ flux and the N2 method, respectively.
\end{abstract}





\section{Introduction}

A Tunable Filter (TF) is an imaging device that can isolate an arbitrary spectral band $\delta\lambda$ at an arbitrary wavelength $\lambda$ over a broad continuous spectral range. Those filters are ideally suited for surveys of emission-line galaxies (ELGs) in different environments and are a powerful tool to detect distant line emitters \citep{Steidel00,Lowenthal91,Macchetto93,Thompson95}. The use of TFs reduces significantly the sky contamination, an important limitation of broad band surveys, because they cover a small wavelength range, thus increasing the contrast between the emission lines and the continuum, and allowing a moderate 2D coverage in a single pointing depending on the instrument. Also, TFs are narrower than most narrow band filters generally used, thus increasing emission line object detection ratio.

Among the first TF systems for non solar astronomy, we have the Goddard Fabry-Perot Imager  \citep[GFPI;][]{Gelderman95}, which is an optical scanning interferometer and CCD imaging system. Also,  \citet{Thompson95} developed a narrowband imaging survey using a Fabry-Perot imaging interferometer. Another known TF system is the Taurus Tunable Filter \citep[TTF;][]{Bland-Hawthorn98a,Bland-Hawthorn98b,Bland03}. Now decommissioned, it was in operation from 1996 to 2003 on the 3.9m Anglo-Australian Telescope and from 1996 to 1999 on the 4.2m William Hershel Telescope \citep[WHT;][]{Bland03}. The TTF was used, among other things, for several extragalactic surveys \citep[e.g.][]{Jones01}, and has shown that there is a rich field of science awaiting exploration with large ground--based telescopes equipped with these narrow--band imagers \citep[for a review see][]{Veilleux05}. Among other instruments with TFs, we have the Maryland-Magellan Tunable Filter \citep[MMTF;][]{Veilleux10}, installed on the Magellan--Baade 6.5m telescope,  located at Las Campanas Observatory, Chile, and the Robert Stobie Spectrograph \citep[RSS;][]{Rangwala08} for the 11m South African Large Telescope (SALT), which provides spectroscopic imaging at any desired wavelength from 430 to 860 nm.

OTELO \citep[OSIRIS Tunable Emission Line Object survey;][]{Cepa08, Cepa05a, Cepa07} is an emission line object survey using the red TF of OSIRIS (Optical System for Imaging and low Resolution Integrated Spectroscopy) \citep{Cepa03,Cepa05b}. The possibility to measure {H$\alpha$} and [{N\,\textsc{ii}}]$\lambda$6583 lines and discriminating AGNs, makes OTELO a unique emission line survey. Observing in selected atmospheric windows relatively free of sky emission lines, it is expected to reach\ a 5 sigma depth of 10$^{-18}$ erg/cm$^2$/s, detecting objects with EW $<$ 0.3 {\AA}. With a total survey area of 0.1 square degrees distributed in different fields, such as the Extended Groth strip, Goods-N, SXDS, and Cosmos, OTELO will be the deepest emission line survey to date. The expected number of emitters distributes as follows: 1000 {H$\alpha$} star forming emitters up to a redshift 0.4, from which about 100 would correspond to low luminosity star forming galaxies, 6000 star forming emitters in other optical emission lines up to a redshift 1.5, 400 Ly$\alpha$ emitters at redshifts up to 6.7, 400 QSO at different redshifts, and about 1000 AGNs. The OTELO survey observations are being presently carried out in the Groth field. The project has produced previous $BVRI$ broad-band photometry \citep{Cepa08}, as well as optical properties of x-ray emitters \citep{Povic09} on this field.

One of the aims of the OTELO survey is to estimate metallicities of ELGs. Among the different indirect methods to estimate metallicities in ELGs we can distinguish between theoretical models, such as [{N\,\textsc{ii}}] $\lambda$6583/[{O\,\textsc{ii}}] $\lambda$3727 \citep{Kewley02}, empirical calibrations, for example the $R_{23}$ ratio \citep{Pilyugin01, Pilyugin05, Liang07}, or a combination of both, e.g. the N2 method \citep{Denicolo02}. A detailed description of the different metallicity methods and calibrations are given in \citet{Kewley08} and  \citet{Lara09a,Lara09b}.

The N2 $\equiv$ [{N\,\textsc{ii}}] $\lambda$6583/ {H$\alpha$} method has been used and calibrated by several authors \citep{Denicolo02,Kewley02,Pettini04,Erb06,Kewley08}, and has demonstrated to work accurately estimating metallicities from 1/50th to twice the solar value \citep{Denicolo02}. One of the most important advantages of this method is that an extinction correction is not required because it only uses the {H$\alpha$} and [{N\,\textsc{ii}}]$\lambda$6583 lines, both close in wavelength. It requires only a narrow spectral range, making it suitable for surveys of limited spectral coverage like OTELO. Finally, the N2 method has demonstrated to work accurately at high redshift ($z \sim$ 2.2), making it suitable for detecting evolution \citep{Erb06}.

The work presented here has been developed within the OTELO project. As part of the OTELO preparatory activities, the aims of this study is to determine the optimal sampling and FWHM combination for the OSIRIS TFs that allows to determine the N2 ratio, by analyzing and recovering flux errors of the {H$\alpha$} and [{N\,\textsc{ii}}]$\lambda$6583 lines. With the selected instrumental configuration it will be possible to deblend both lines, classify galaxies as star--forming and AGNs using the N2 ratio \citep{Stasinska06}, to estimate the star formation rates (SFRs) with the {H$\alpha$} flux \citep[e.g.][]{Kennicutt98}, and calculating the chemical abundances using the N2 method \citep{Denicolo02} in star forming galaxies.

This paper is structured as follows: in Sect. 2 we give a review of the OSIRIS instrument, in Sect. 3 we detail the scanning tunable imaging technique, in Sect. 4 we analyze the error estimates, in Sect. 5 we try our method using SDSS data, and finally, in Sect. 5 we give the conclusions.

\section{OSIRIS's Tunable Filters}

OSIRIS  is the Spanish Day One instrument for the GTC 10.4m telescope. With a field of view of 8.5x8.5 arcminutes and sensitive in the wavelength range from 3650 through 10000 ${\AA}$, OSIRIS is a multiple purpose instrument for imaging and low-resolution long slit and multiple object spectroscopy (MOS). The main characteristic of OSIRIS is the use of two TFs, one for the blue (3700-6700 {$\AA$}), and another for the red (6400-9600 {$\AA$}) that overlap in wavelength and allow to cover most of the full OSIRIS wavelength range \citep{Cepa08}.

Tunable narrow-band filters, also known as Fabry-Perot filters (FPFs), consist essentially of two glass or quartz parallel plates with flat surfaces enclosing a plane--parallel plate of air. The inner surfaces are coated with films of high reflectivity and low absorption.

The general equation for the intensity transmission coefficient of an ideal FPF (an Airy Function), as a function of wavelength is:

\begin{equation}
\tau_{r}(\lambda)=\left({\frac{T}{1-R}}\right)^2\left[1+{\frac{4R}{(1-R)^2}}sin^2\left({\frac{2{\pi}{\mu}dcos{\theta}}{\lambda}}\right)\right]^{-1}
\end{equation} where $T$ is the transmission coefficient of each coating, $R$ is the reflection coefficient, $d$ is the plate separation, ${\mu}$ is the refractive index of the medium in the cavity, usually air with ${\mu}=1$, and ${\theta}$ is the angle of incidence.


The instrumental response of an ideal FPF, given by Equation 1, is periodic in wavelength and formed by Airy profiles, as show in Fig. 1.  See \citet{Bland89} and  \citet{Born99} for a detailed theory explanation.


\begin{figure*}[ht!]
\centering
\includegraphics[scale=0.60]{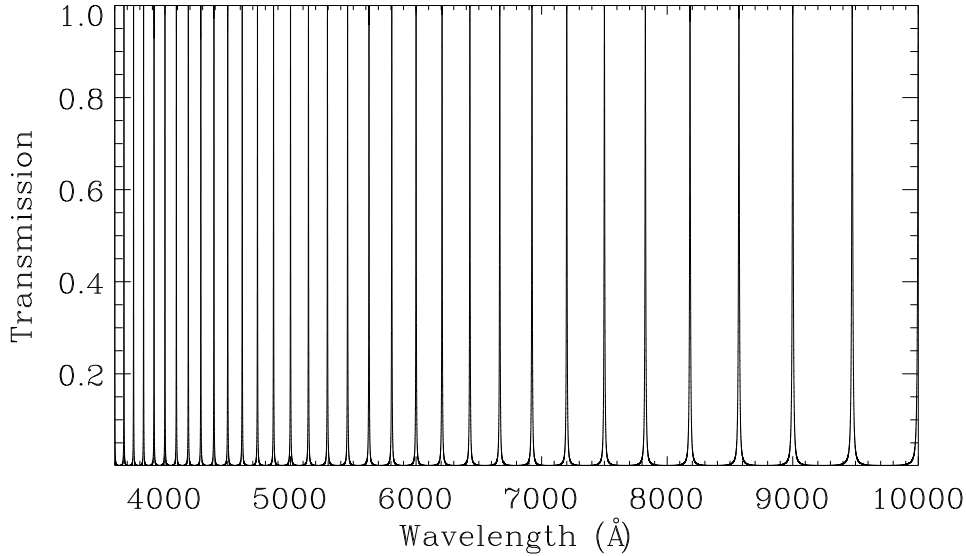}
\caption{Transmission of a tunable filter as a function of wavelength.}
\end{figure*}

According to the OSIRIS characteristics, the available TF FWHM as function of wavelength span in a range from $\sim$8 to $\sim$20 {\AA}. \footnote{http://www.gtc.iac.es/en/pages/instrumentation/osiris/data-commissioning.php$\#$OSIRIS$\_$TF$\_$filter$\_$widths}

 \section{Scanning tunable imaging technique}

We can define scanning tunable imaging as taking a set of images of the same FOV with the TF tuned at different contiguous wavelengths, which is similar to low resolution MOS spectroscopy. Each wavelength is shifted by a certain fraction of the TF FWHM with respect to the others \citep[e.g.][]{Jones01,Cepa10}.

As part of the preparatory activities for OTELO, we simulate the scanning using a tunable filter with different FWHM of the spectra of several {H\,\textsc{ii}} galaxies, aimed at selecting the best combination of tunable filter FWHM and sampling. This combination will allow deblending H$\alpha$ from [{N\,\textsc{ii}}]$\lambda$6583 lines with a flux relative error lower than 20$\%$ (5$\sigma$ error), which is the maximum error for reliable sources and flux emission-line detection according to the project requirements \citep[see also][]{LaraLopez10}.

Given the low FWHM of the TF, it will be possible to estimate the object's chemical abundances using the N2 method even for very low metallicity systems.

As a first step, we generated the response of the TF (an Airy function), with Equation 1, observing that a difference of at least 3 {\AA} in FWHM is required for obtaining significant differences in the recovered lines-flux error. Then we perform several tests with  FWHMs of 6, 9, 12, and 15 {\AA}. However, according to the characteristics of OSIRIS as explained in Sect. 2, a FWHM of 6 {\AA} is not available, and a FWHM of 15 {\AA} gives errors larger than 25$\%$ for  [{N\,\textsc{ii}}]/H$\alpha$, which is out of our upper limit error. Therefore, we selected FWHMs of 9 and 12 ${\AA}$, as shown in Fig 2.


\begin{figure}[ht!]
\centering
\includegraphics[scale=0.4]{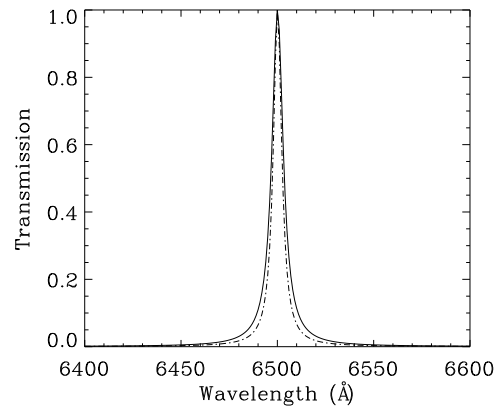}
\caption{An Airy function with FWHM of 9 and 12 ${\AA}$, in solid and dot-dash lines, respectively.}
\end{figure}

\subsection{Generation of synthetic spectra}

We generated synthetic spectra based on data from real {H\,\textsc{ii}} galaxies, with {H$\alpha$} and [{N\,\textsc{ii}}]$\lambda$6583 lines in emission centered on 6563 and 6583 {\AA} respectively, FWHM({H$\alpha$})=4.7 {\AA}, and [{N\,\textsc{ii}}]$\lambda$6583/{H$\alpha$}=0.43, which correspond to a maximum rotation velocity ($V_{max}$) of 215 km/s. Median values of $V_{max}$ decrease from 300 to 220 to 175 km/s for the Sa, Sb, and Sc types, respectively \citep{Roberts78, Rubin85, Sandage00,Sofue01}, then, our synthetic spectra are representative of spiral galaxies.

We redshifted the spectra to z=0.24 and z=0.4, the two redshifts of the chosen atmospheric windows of 150 and 180 {\AA} width, respectively. The wavelengths at z=0.24 and z=0.4 are of 8138 \& 9188 {\AA} for {H$\alpha$}, respectively, and of 8163 \& 9216 {\AA} for [{N\,\textsc{ii}}]$\lambda$6583, respectively. At redshift zero, {H$\alpha$} and [{N\,\textsc{ii}}]$\lambda$6583 lines are separated by $\sim$20{\AA}. As redshift increases, the separation between both lines increases as 1+$z$ (25 {\AA} at z=0.24 and 28 {\AA} at z=0.40), making it easier to deblend {H$\alpha$} from [{N\,\textsc{ii}}].\\

The intermediate observed redshift populations at 0.24 and 0.4 are representative of the transition from the relative quiet local universe to the starbursting universe at  z$\sim$1. For instance, galaxies at redshift 0.4 have shown lower metallicity and higher SFR than those of the local universe \citep{Lara09a,Lara09b,Lara10a}.

To add a continuum, we used H$\alpha$ EWs of 5, 10, 20, 30, 40 and 50 {\AA}, with the EW defined by  EW$_{{\rm H}\alpha} = {\frac{F_{{\rm H}\alpha}}{F_{c,{\rm H}\alpha}}}$, where $F_{c,{\rm H}\alpha}$ is the continuum flux density at the H$\alpha$ line, and $F_{{\rm H}\alpha}$ is the {H$\alpha$} flux  of the ELG \citep{Waller90}. The adopted EWs assure the inclusion of several morphologies and types of galaxies \citep{Kennicutt83,Kennicutt98,Gavazzi06}, as shown in Table 1.

Finally, to add noise to the spectra, we considered the equation of the signal--to--noise ($S/N$) of a charge-coupled device (CCD), or the $``$CCD Equation$"$ \citep{Mortara81,Newberry91,Gullixson92}:

\begin{equation}
S/N=\large{{\frac{N_{\ast}}{\sqrt{ N_{\ast}+n_{pix}(N_S+N_D+{N_R}^2)}}}}
\end{equation} where $N_{\ast}$ is the total number of photons (signal), $n_{pix}$ is the number of pixels under consideration for the $S/N$ calculation, $N_S$ is the total number of photons per pixel from the background or sky, $N_D$ is the total number of dark current electrons per pixel, and ${N_R}^2$ is the total number of electrons per pixel resulting from the read out noise.

We can see from Equation 2 that if the total noise for a given measurement is dominated by the first noise term, $N_{\ast}$, the Equation 2 becomes $S/N=\sqrt{N_{\ast}}$, which is a measurement of a single Poisson behaved value. Therefore, we add a Poisson noise to the simulated spectra. We adopted a $S/N$ of 5, which ensures the detection of the object within an error of $\pm20\%$. The magnitude error of the observed object is $\Delta$Mag $\simeq N/S$, which means a 0.2 magnitude error for a $S/N$ of 5. This procedure is valid when sky noise is not dominant.

\begin{table}[h]
\begin{center}
\begin{tabular}{c|c}   
\hline
Type & EW(H$\alpha$) ({\AA})\\\hline\hline
E& 0\\\hline
Sab& 2-40\\\hline
Scd/Im& 10-100\\\hline
HII/BCD& 20-400\\\hline
Sy2/Sy1& 86-260\\\hline
\end{tabular}
\end{center}
\caption{Equivalent widths of galaxies with different morphological types. E, Sab, and Scd/Im from \citet{Kennicutt83}, \citet{Kennicutt98}, \citet{Gavazzi06}; {H\,\textsc{ii}}/BCD from \citet{Gil03}; and Sy2/Sy1 from \citet{Gallego97}.}
\end{table}

\subsection{Convolutions}

According to the convolution theorem, convolution in one domain equals point-wise multiplication in the other domain (e.g., frequency domain), thus we multiply the functions we want to convolve, the Airy function, and the HII galaxy spectra. For the point to point multiplication it is important that both functions have the same resolution.

We take into account the following variables for the convolutions:

   \[
      \begin{array}{lp{0.8\linewidth}}

& -FWHM of the Airy function (9 $\&$ 12 ${\AA}$)\\
& -Sampling (1, 2, 3, 4, 5, 6, 7, 8, 9, $\&$ 10 ${\AA}$)\\
& -EW of the spectra (5, 10, 20, 30, 40 $\&$ 50 {\AA})\\
& -Initial wavelength for sampling (8075-8084 for z=0.24 and 9110-9119 for z=0.4).\\
      \end{array}
   \]

The combination of all these variables allow exploring a fairly complete set of possibilities.
We then convolved the FWHM of the Airy function with the different EW--spectra according to the following procedure, where $n$ is the sampling, and $i$ is a counter that goes from 0 to 149/$n$ for $z=0.24$, and from 0 to 179/$n$ for $z=0.4$ (which are the two redshifts where the atmospheric windows are located), following this procedure:

 \begin{enumerate}
\item We set the peak of the Airy function at an initial wavelength $\lambda_0+ni$ of the spectrum. We start at $\lambda_0$ for $i=0$.
\item We convolve both functions and integrate the resultant flux in a fixed window of 150 ${\AA}$ for the spectra at z=0.24, and 180 ${\AA}$ for z=0.4 (i.e. similar to the spectral range of the sky windows considered).
\item We continued shifting the peak of the Airy function $n$ ${\AA}$ to $\lambda_0+n$ (for $i=1$), and convolve again the Airy function with the spectrum and so on up to $i=149/n$ and $i=179/n$ for  $z=0.24$ and 0.4, respectively, as show in Fig 3.
\item The integrated fluxes are plotted versus wavelength $\lambda_0+ni$ for generating a pseudo--spectrum.
 \end{enumerate}

In this way, we construct several pseudo--spectra, some of them shown in Figs. 4 and 5 for FWHM 9 and 12 {\AA}, respectively, and an EW({H$\alpha$}) of the original spectrum of 50 {\AA} in both cases. Each point in those figures represents the integrated flux after every convolution as would be obtained from aperture photometry on the images. Using all the combinations of FWHM of the Airy function, the spectra of several EW (5, 10, 20, 30, 40 and 50 {\AA}), and different samplings (from 1 to 10 {\AA}), we obtained a total of 120 pseudo--spectra, one for every combination of FWHM, sampling, and spectra.

Although a pseudo--spectrum looks like a spectrum, we should emphasize that it is not, since every point represented corresponds to the integrated flux resulting from the convolution of the spectrum with the response of the tunable filter, in a discrete, non continuous way.

\begin{figure}[h]
\centering
\includegraphics[scale=0.3]{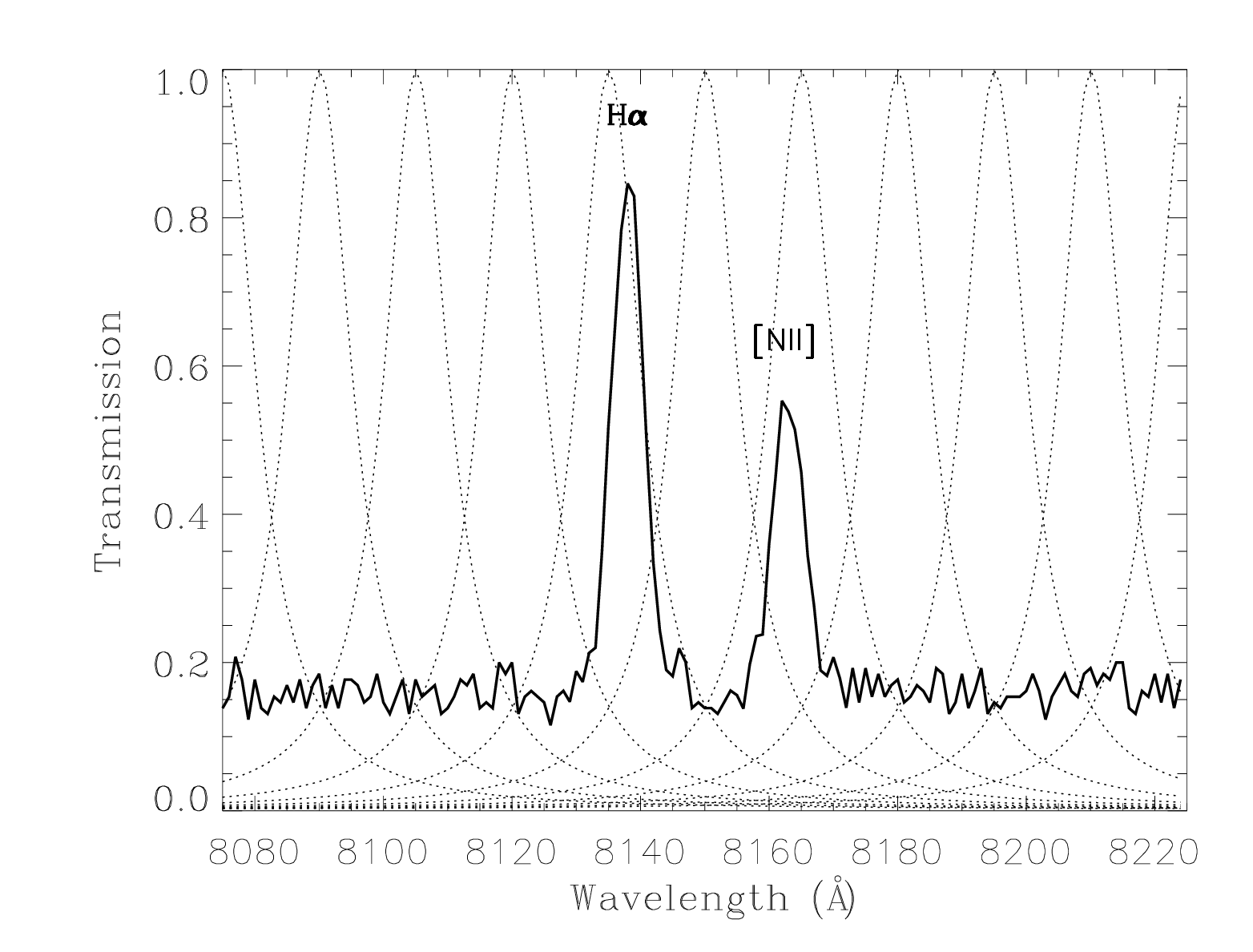}
\caption{Example of a spectra with {H$\alpha$} and [{N\,\textsc{ii}}]$\lambda$6583 lines in emission with a $S/N$ of 5, sampled every 10 {\AA} by an Airy function of 12 ${\AA}$ FWHM.}
\end{figure}

\begin{figure*}[t!]
\centering
\includegraphics[scale=0.29]{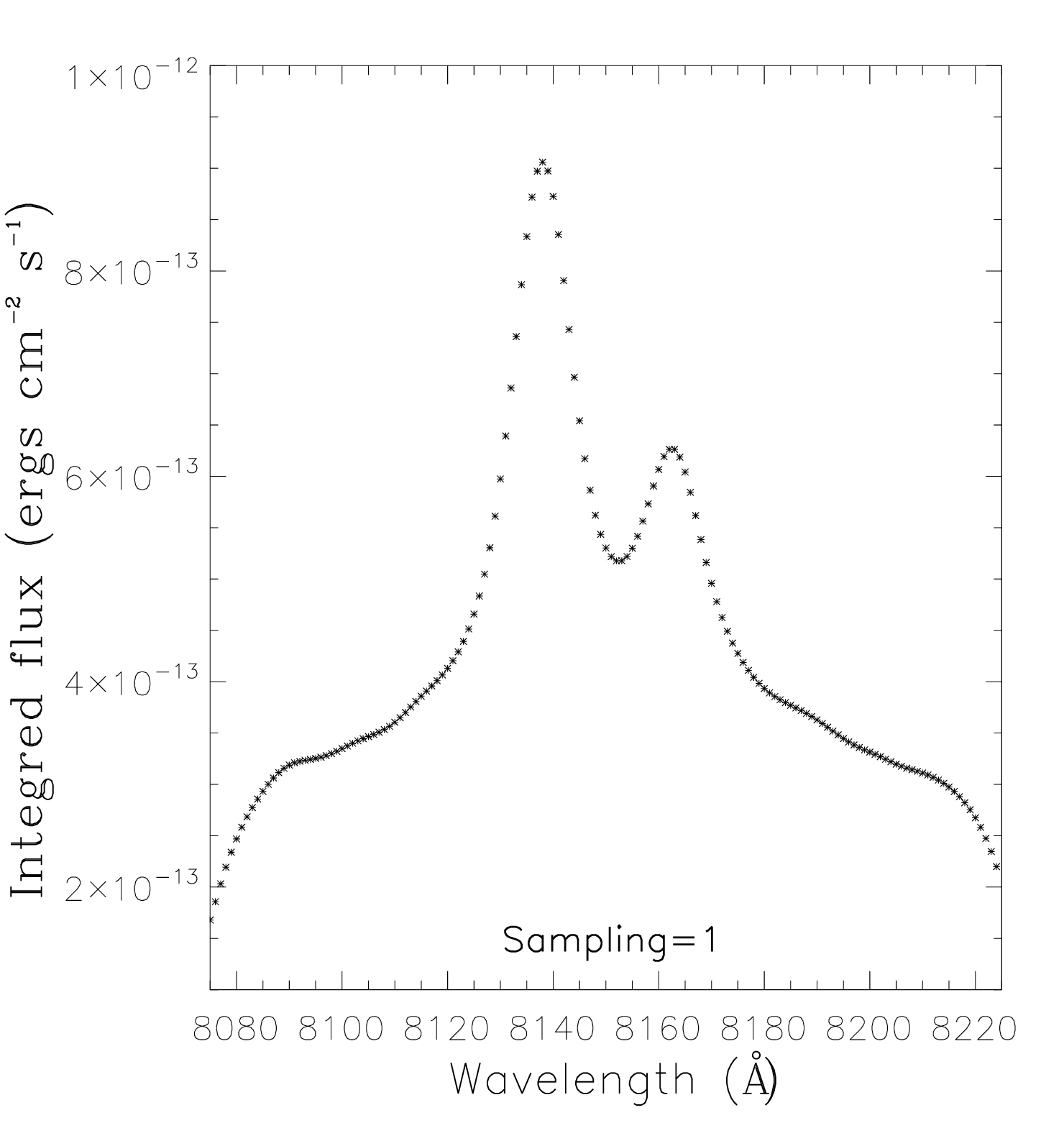}
\includegraphics[scale=0.29]{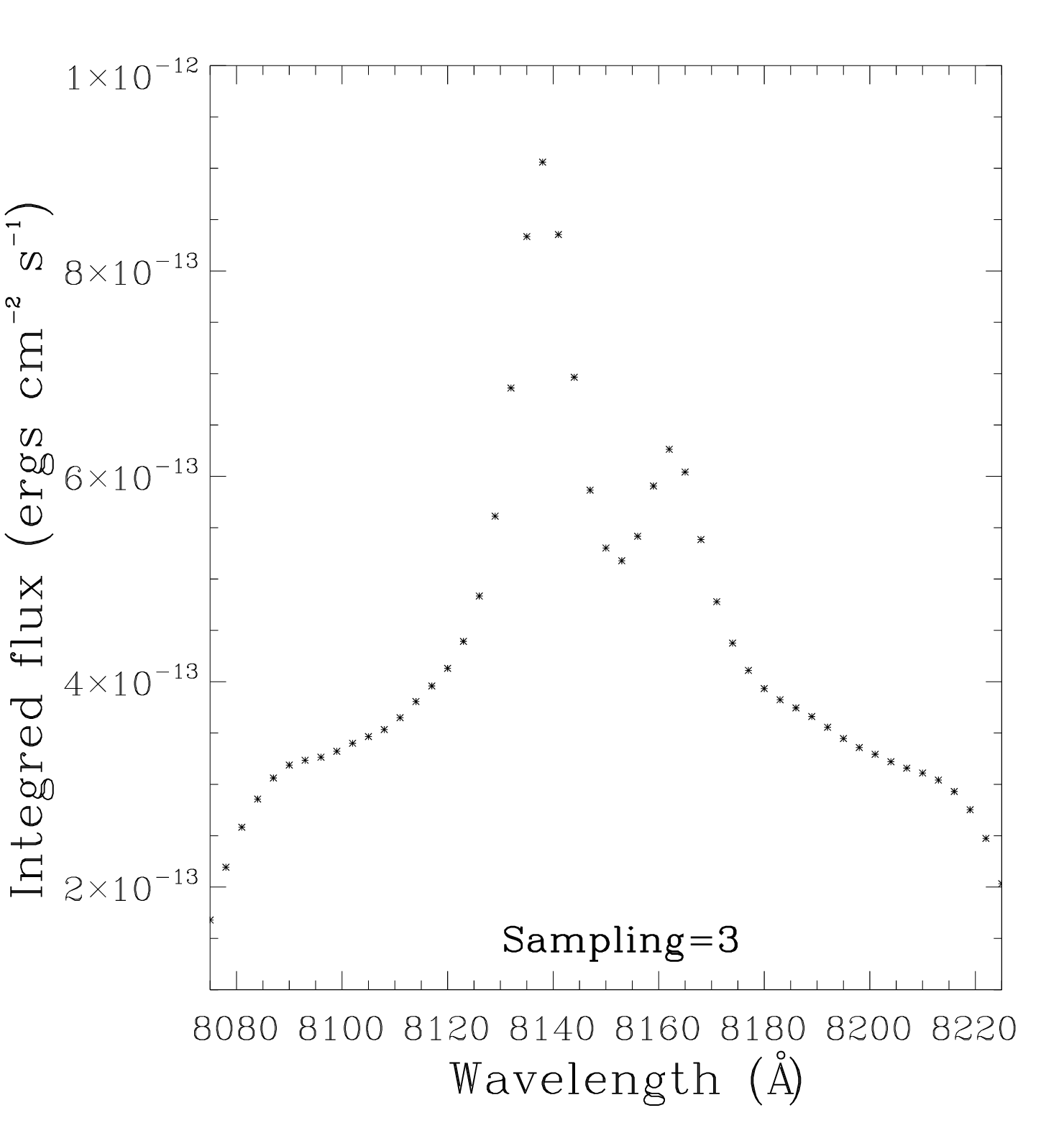}
\includegraphics[scale=0.29]{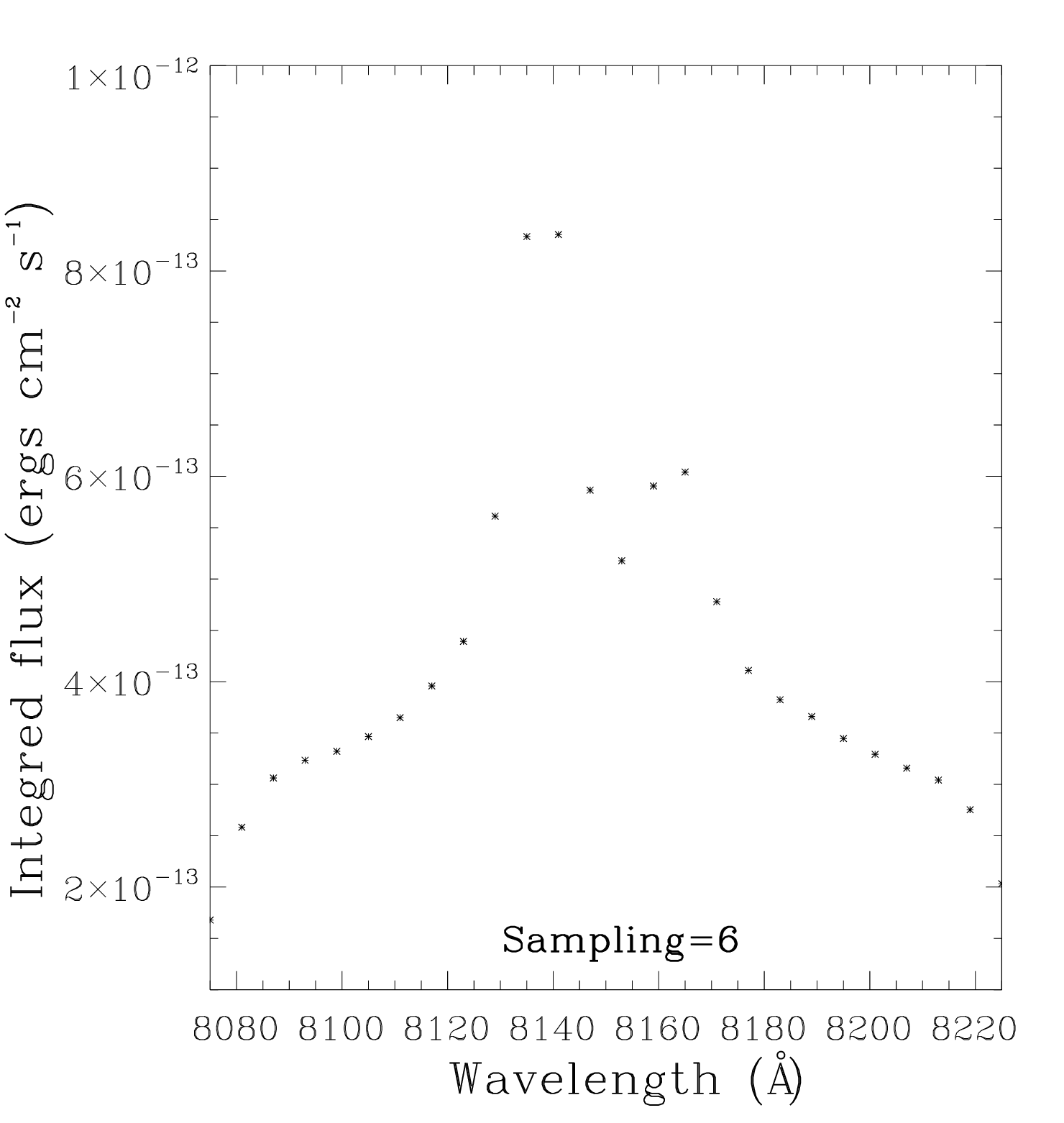}

\caption{Pseudo--spectra resulting from the convolution of a spectrum at $z=0.24$ of an EW of 50 ${\AA}$ with an Airy function of FWHM of 12 ${\AA}$, sampling every 1, 3, and 6 ${\AA}$.}
\end{figure*}

\begin{figure*}[t!]
\centering
\includegraphics[scale=0.29]{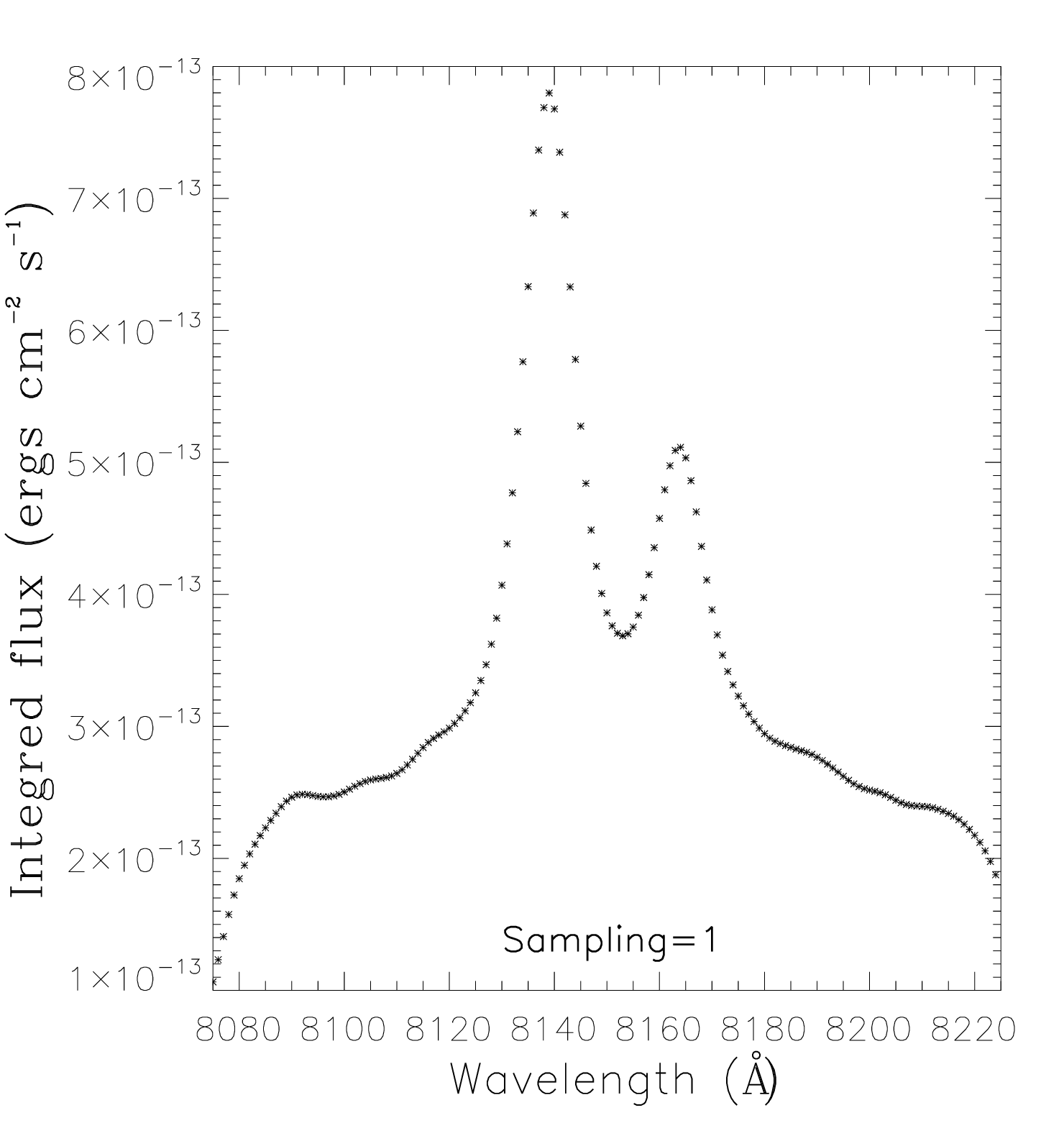}
\includegraphics[scale=0.29]{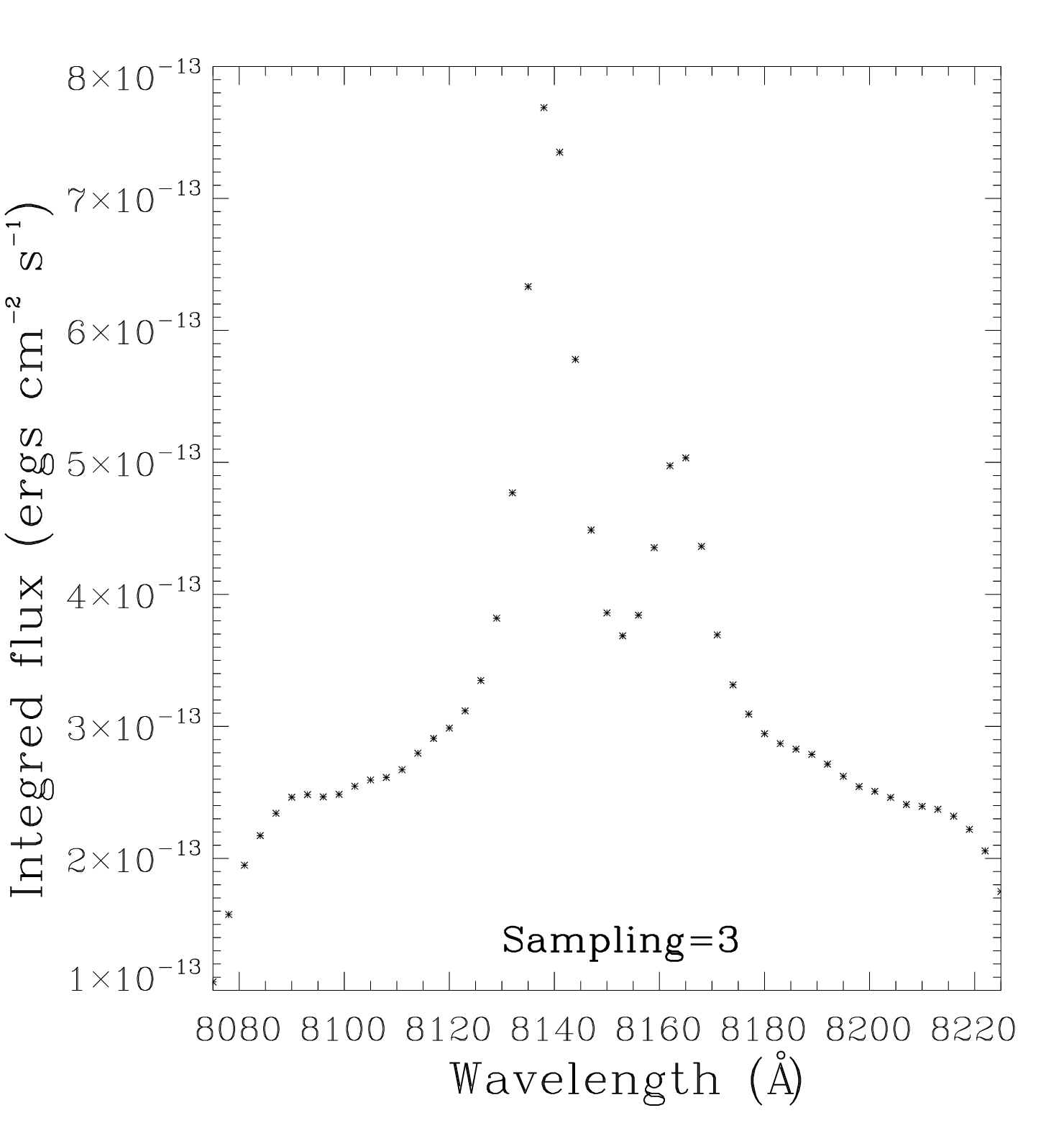}
\includegraphics[scale=0.29]{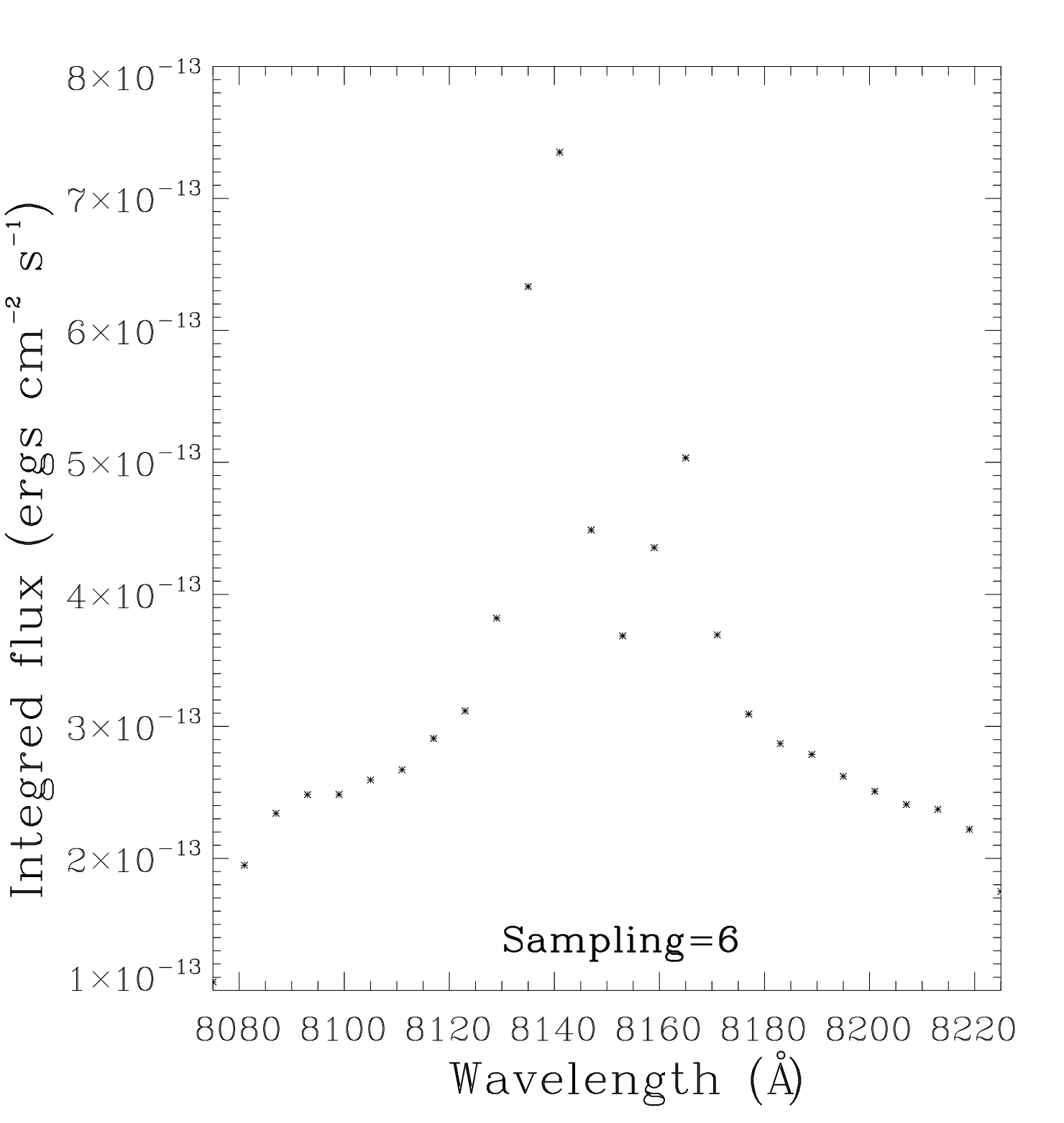}

\caption{Pseudo--spectra resulting from the convolution of a spectrum at $z=0.24$ of an EW of 50 ${\AA}$ with an Airy function of FWHM of 9 ${\AA}$, sampling every 1, 3, and 6 ${\AA}$.}
\end{figure*}

In Figs. 4 and 5 we notice a drop off at the edges of the pseudo--spectra. This is due to the limits on the wavelength integrated interval chosen. In a real case, the spectral range is limited by the order sorter used, and the operating wavelength range is lower than the FWHM of the order sorter and then this effect is barely noticed.


\subsection{Continuum subtracted and flux estimates}

Before estimating the flux error of the emission lines, we subtract the continuum from the pseudo--spectrum. As a first approximation to subtract the continuum, we fitted an horizontal line to the pseudo--spectrum continuum, and estimated the {H$\alpha$} and [{N\,\textsc{ii}}]$\lambda$6583 line fluxes, but this procedure results in large errors. We find that a better method is to fit the continuum of the pseudo--spectrum as shown in Fig. 6 (left). We proceed to simulate a spectrum with the same characteristics of S/N, but without any emission lines, and proceed to convolve it as described above. In this way, we obtained a pseudo--spectrum of the continuum. It is important to notice that the best method to subtract the continuum when dealing with real observations, will be to fit a function with the form of the entire pseudo--spectrum.

If the FWHM of the observed line is of the same size or larger (i.e. quasars) than that of the Airy function, then it is possible to recover the flux and FWHM of the line through a deconvolution. However, if the FWHM of the observed line is smaller than that of the Airy function, as it is usually the case, a deconvolution is not useful for recovering the fluxes or FWHM of the observed lines, as we found in a first test. Nevertheless, we observed that the peak corresponding to the H$\alpha$ and [{N\,\textsc{ii}}]$\lambda$6583 lines in the pseudo--spectrum is enough for recovering the fluxes because it has the information of the integration of the entire line. Therefore, from the continuum subtracted pseudo--spectrum (see Fig. 6, right), the H$\alpha$ and [{N\,\textsc{ii}}]$\lambda$6583 fluxes were estimated from the corresponding peak of each line in the pseudo--spectrum. This is clearly one of the main differences with respect to spectroscopic data.

\begin{figure*}[th!]
\centering
\includegraphics[scale=0.3]{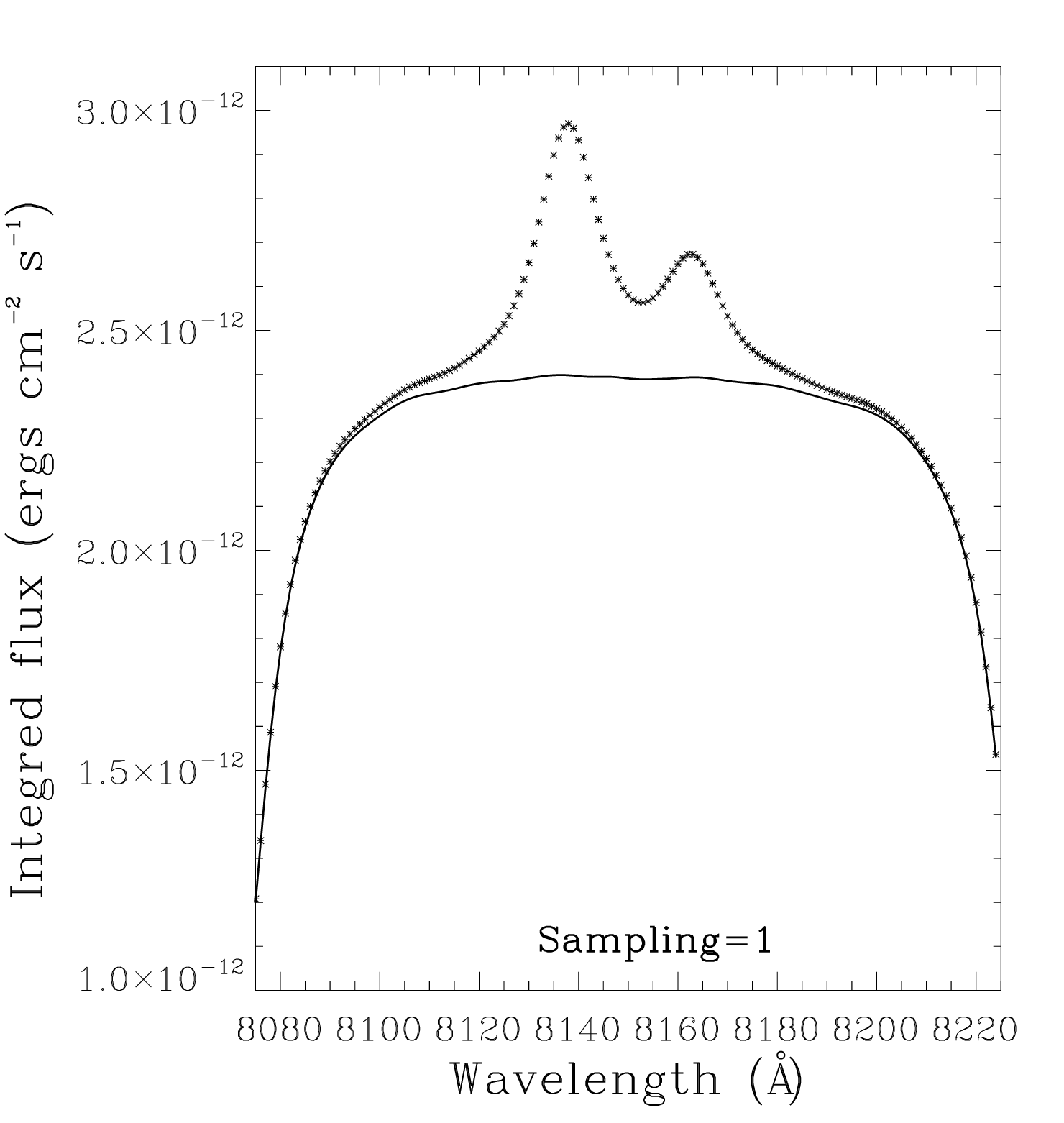}
\includegraphics[scale=0.3]{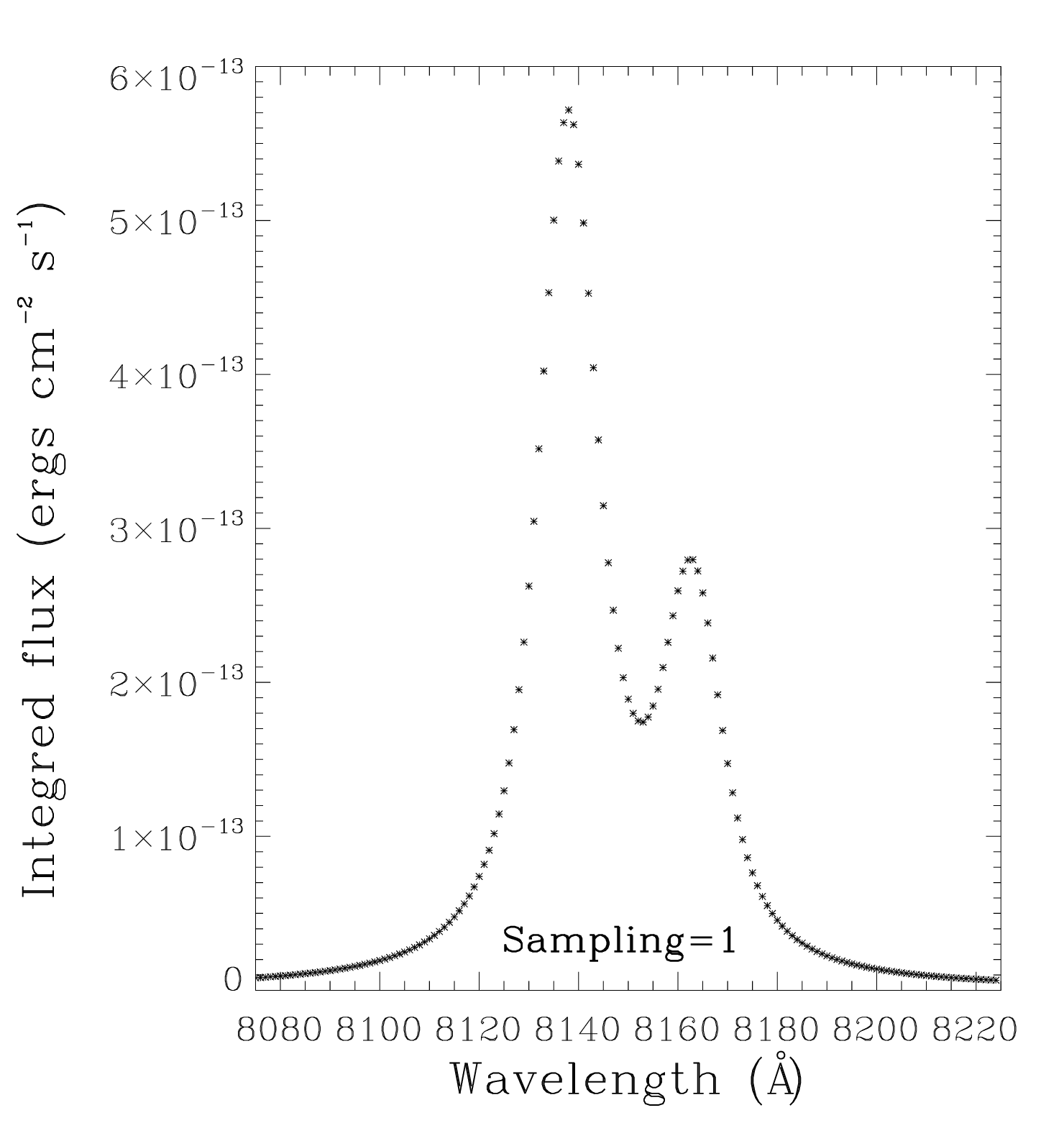}
\caption{Left: Pseudo--spectum resulting from the convolution of a spectrum with an EW of 5 ${\AA}$ with an Airy function with FWHM of 12 ${\AA}$, sampling each 1 ${\AA}$, and fitting the continuum with a solid line. Right: Result of subtracting the continuum to the pseudo--spectum.}
\end{figure*}

\section{Error estimation}
\label{sec:5}

In order to obtain the best combination of TF FWHM and sampling that allow deblending H$\alpha$ from [{N\,\textsc{ii}}]$\lambda$6583, we obtained relative errors from the recovered fluxes for all the combinations of TF FWHM, sampling, redshifts, and spectra EWs. One of the principal requirements for selecting the optimal combination of TF FWHM and sampling, will be to obtain a line flux error lower than 20$\%$. This error will assure an ELG detection, and a reliable line flux.

The H$\alpha$ and [{N\,\textsc{ii}}]$\lambda$6583 fluxes were obtained from the peaks of the pseudo--spectrum corresponding to each line, as explained above. The contamination from the nearby lines (H$\alpha$ or  [{N\,\textsc{ii}}]$\lambda$6583), will depend on the FWHM of the employed Airy function and most of the sampling interval. A large FWHM of the Airy function will certainly enclose a high percentage of the flux of the emission line when TF line FWHM are comparable. However, it will also cause a higher percentage of contamination from closer lines. On the contrary, a small FWHM of the Airy function will result in higher errors recovering the emission line flux depending on their widths, but also in a smaller contamination from closer lines. Therefore, the analysis of the error estimates will  allow us to obtain the best TF FWHM that better recovers the original flux line with the least contamination from other lines. We estimated relative errors (defined as the value of absolute difference between measurement and the real value, divided by the real value) from the comparison of the recovered emission-lines of the pseudo--spectra with the original lines fluxes of the simulated spectra.

\begin{figure*}[t!]
\centering
\includegraphics[scale=0.6]{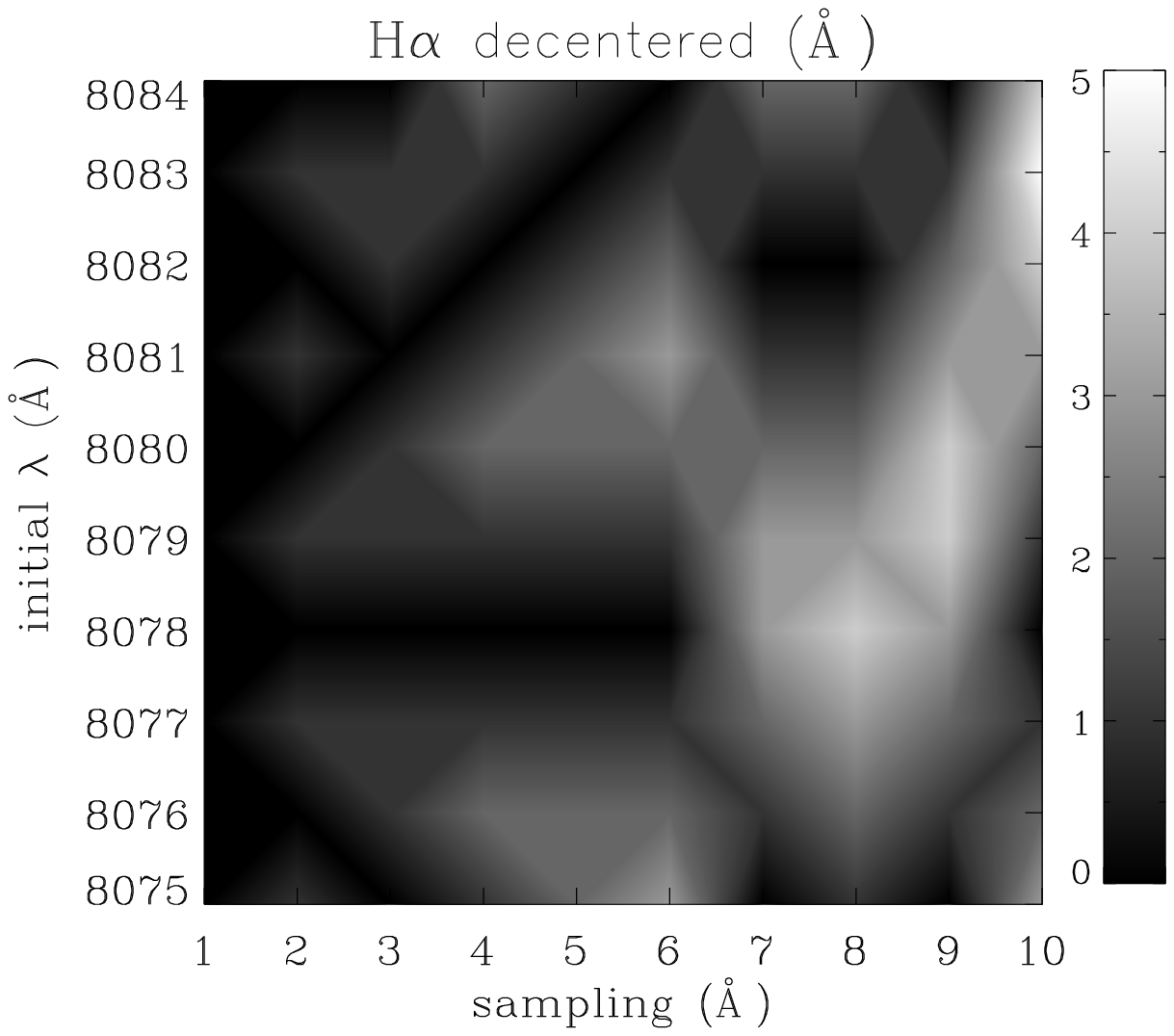}
\includegraphics[scale=0.6]{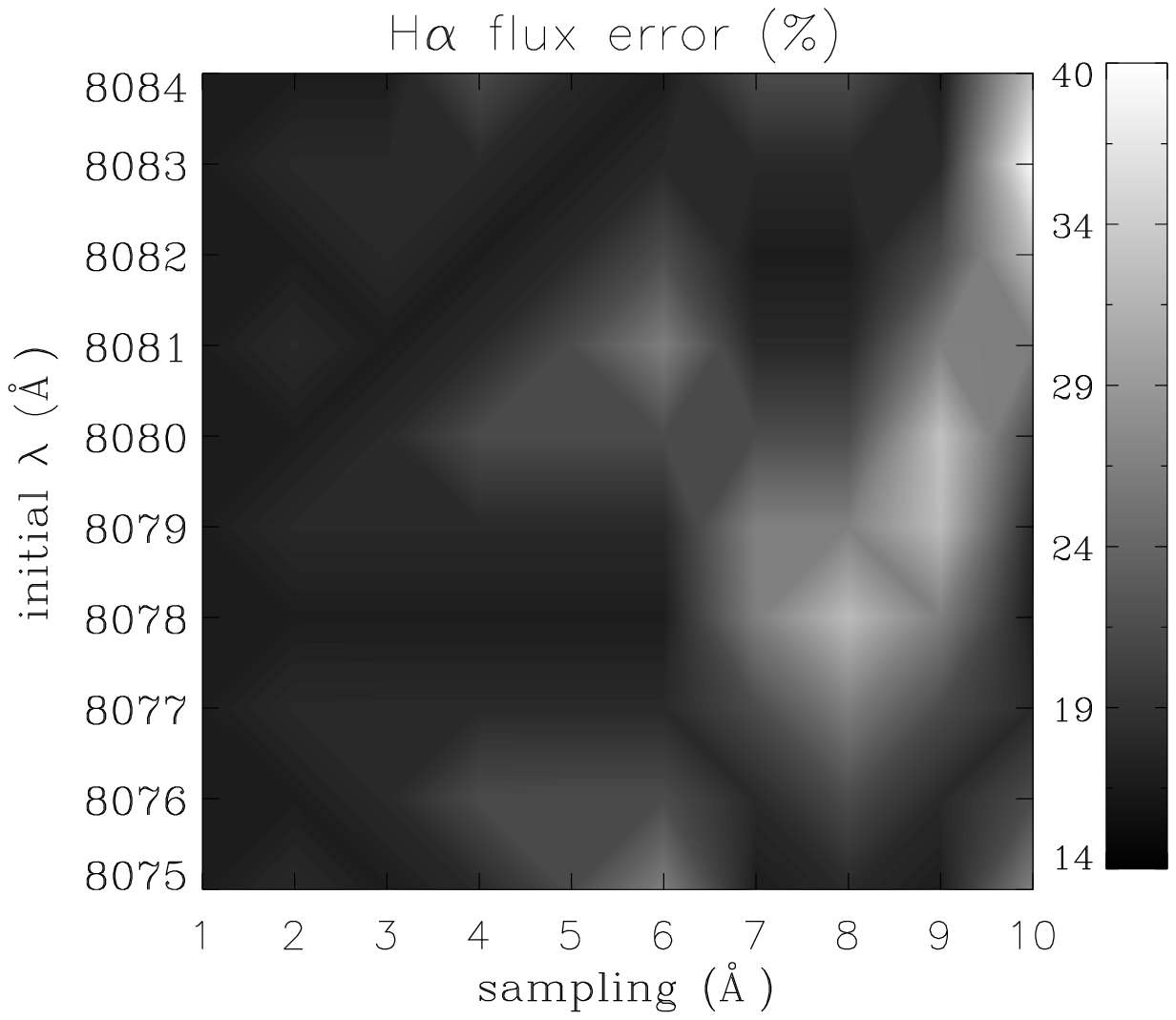}
\caption{Density plots of the errors obtained using a spectrum with EW(H$\alpha$)=50{\AA} and $z=0.24$, convolved with an Airy function of FWHM=12{\AA} as a function of sampling and starting $\lambda$. Left: errors in the H$\alpha$ decentered ({\AA}). Right: H$\alpha$ flux error ($\%$)}
\end{figure*}

\begin{figure*}[ht!]
\centering
\includegraphics[scale=0.5]{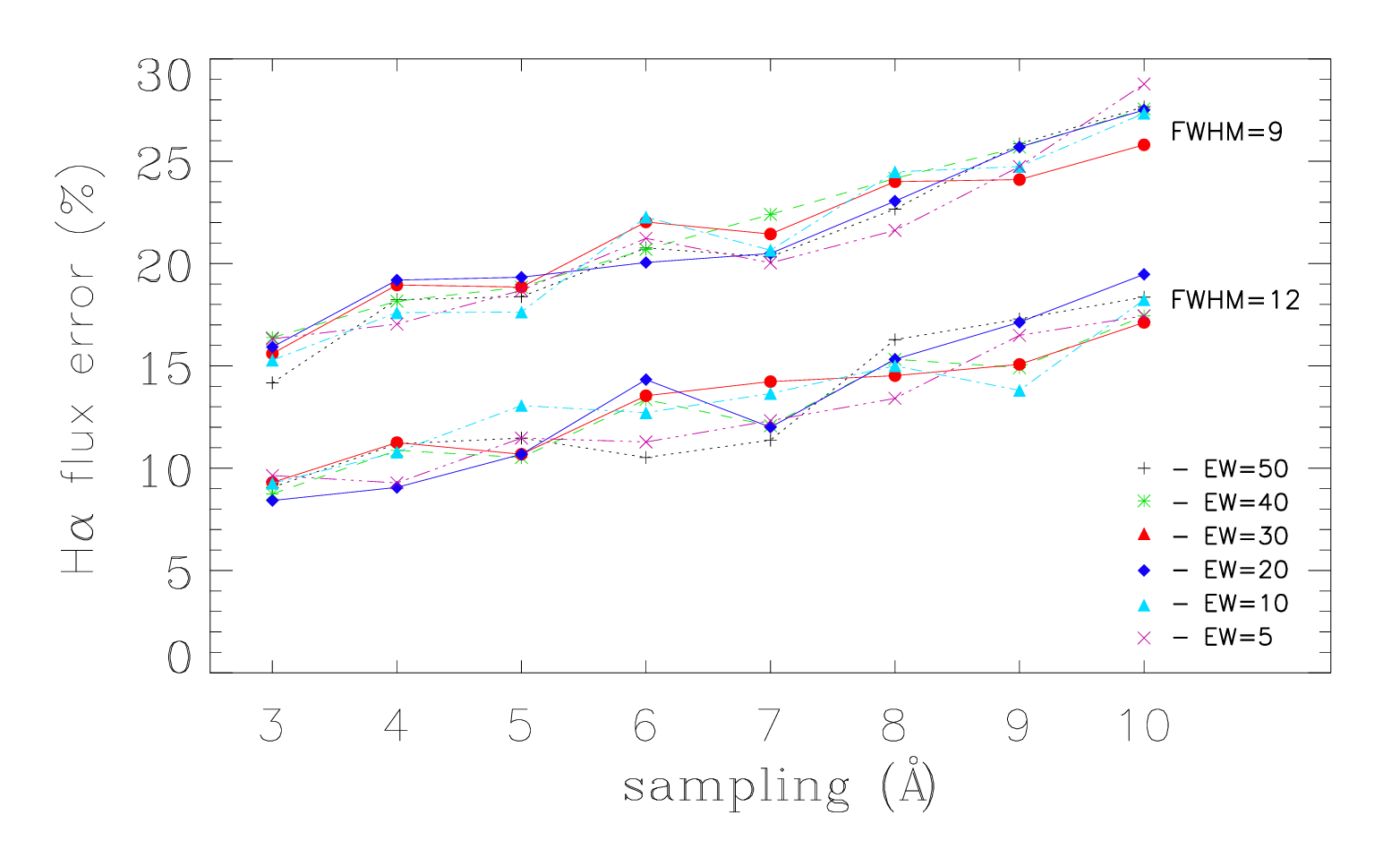}
\caption{H$\alpha$ flux errors obtained using a spectrum with EW(H$\alpha$) of 50, 40, 30, 20, 10, and 5 {\AA}, convolved with an Airy function of FWHM of 12 and 9 {\AA}, and sampling from 3-10 {\AA}. Data taken from Tables 6 to 9. Symbols are indicated in the Figure.}
\end{figure*}

We have also analyzed the wavelength errors of the two peaks of the pseudo--spectra that would correspond to the H$\alpha$ and [{N\,\textsc{ii}}]$\lambda$6583 lines. Ideally, the peak of the pseudo--spectra would indicate the original emission line center, but its error will depend on the initial wavelength and the sampling interval. This is an important point to take into account because this peak would be also indicative of the redshift of the detected sources. The highest difference in wavelength of the pseudo--spectra peak (detected line) with the observed one, will be half the sampling interval. For example, the error in the emission line center of a detected source convolved using a sampling=6${\AA}$, will be of $\pm3{\AA}$. However, when fitting a profile to the pseudo--spectra, the error in the emission line center would decrease.

In Fig. 7 we show density plots of the errors obtained using a spectrum with EW(H$\alpha$)=50 {\AA} and z=0.24, convolved with an Airy function of FWHM=12${\AA}$ as function of sampling and starting {$\lambda$}. In the left panel the difference in wavelength of the original center of the spectral line with respect to that obtained from the pseudo--spectrum is shown. The right panel shows the relative flux error of the  H$\alpha$ line. In both panels we use a sampling from 1 to 10 ${\AA}$ and ten consecutive initial wavelengths. Both figures show the same patterns, with large decentering errors (Fig. 7 left) producing large errors in the recovered flux (Fig. 7 right). Although a sampling lower than $\sim$ 3 {\AA} would be not realistic, due to the large observing time needed to complete the scan, it is included in the plots for completeness.

Using each one of the simulated spectra of Sect. 3.1, we obtained the relative errors sampling from 1 to 10 ${\AA}$ at ten different starting wavelengths (to be consistent with the largest sampling), in such a way that for every sampling value we obtained ten different relative errors of the recovered emission lines. We then estimated the median error value of the ten different initial wavelengths for every sampling, as shown in Tables 6 to 9. In those tables we show the relative flux errors of the recovered H$\alpha$ and [{N\,\textsc{ii}}]$\lambda$6583 lines, as well as the error of its ratio, sampling from 3 to 10 ${\AA}$ for the different spectra and FWHM of the Airy function. Smaller samplings are not included because in real observations the observing time would be prohibitive.

In Fig. 8 and Tables 2 to 5 can be appreciated that errors increase with sampling, and as a result their standard deviation as well. The H$\alpha$ and [{N\,\textsc{ii}}]$\lambda$6583 errors corresponding to the FWHM of 9${\AA}$, are higher than those using a FWHM of 12${\AA}$. However, the error of the lines ratio is lower. As sampling interval increases, also do the errors, but the total integration time decreases. Therefore it is important to select the sampling whose errors compensate with the total integration time. Although the error of the line ratios is lower using a FWHM of 9 ${\AA}$, the H$\alpha$ error is higher than that using a FWHM of 12${\AA}$ and, as consequence, the error in the SFRs estimate would be larger. For the OTELO project, a TF FWHM of 12 ${\AA}$ and a sampling of 5 ${\AA}$ have been selected because their errors are lower than 20$\%$ for all the EWs, and they are only slightly higher than those with a sampling of 4 ${\AA}$ (see Fig. 8).

\section{Working with real SDSS data}

In order to test the efficiency of the proposed FWHM bandwith and sampling, we apply our method to some galaxy spectra from the Sloan Digital Sky Survey--Data Release 7 (SDSS --DR7) \citep{York00,Abaza09}. The SDSS spectra were obtained using 3 arcsec diameter fibres with a 2.5 m telescope located at Apache Point Observatory \citep{Gunn06}, covering a wavelength range of 3800-9200 {\AA}, and with a mean spectral resolution $\lambda$/$\Delta\lambda$ $\sim$ 1800. Further technical details can be found in Stoughton et al. (2002).

We selected a total of four galaxies from the SDSS-DR7 of different N2 ratios to test the efficiency of our method, two at $z \sim 0.24$, and two at $z \sim 0.4$. At each redshift we selected a star--forming galaxy (spSpec-52368-0580-499, spSpec-53816-2231-307), and an AGN (spSpec-53491-2097-516, spSpec-53473-2108-507). As observed in Fig. 9 and Tables 2 to 5, we selected ELGs of different  H$\alpha$ and [{N\,\textsc{ii}}]$\lambda$6583 intensities, in some cases both lines have similar intensities (e.g. spSpec-53491-2097-516), and in other the [{N\,\textsc{ii}}]$\lambda$6583 line is weak (e.g. spSpec-53816-2231-307). It can be also appreciated the different morphologies of the SDSS galaxies, including a spiral (spSpec-53491-2097-516), a SO/Sa spiral (spSpec-53473-2108-507), and compact galaxies (spSpec-52368-0580-499, spSpec-53816-2231-307).

Although with emission lines it is not possible to estimate metallicities in AGNs, we have included them because we expect to be able to observe and classify AGNs in the OTELO survey. AGNs can be differentiated from star forming and composite galaxies using the N2 ratio as follows: star forming galaxies those with log([{N\,\textsc{ii}}]/{H$\alpha$}) $\leq -0.4$, composite galaxies those with $-0.4 <$ log([{N\,\textsc{ii}}]/{H$\alpha$}) $\leq -0.2$, and as AGNs those galaxies with log([{N\,\textsc{ii}}]/{H$\alpha$}) $> -0.2$ \citep{Stasinska06}. For details and errors of this classification see also \citet{Lara10a}.

In Fig. 9, we present the image of the galaxies, the section of the SDSS spectra that shows the H$\alpha$ and  [{N\,\textsc{ii}}]$\lambda$6583 lines in emission, and in Tables 2 to 5, we show some information about the galaxy spectrum, such as its redshift and the ratio H$\alpha$/[{N\,\textsc{ii}}]$\lambda$6583, where the emission line fluxes were estimated fitting a gaussian to the original spectra. The center (\AA), height (10$^{-17}$ergs cm$^{-2}$ s$^{-1}$ {\AA}$^{-1}$), sigma (\AA), and flux (10$^{-17}$ergs cm$^{-2}$ s$^{-1}$ {\AA}$^{-1}$) of H$\alpha$ and [{N\,\textsc{ii}}]$\lambda$6583 of the original spectra are also shown. We have convolved those galaxy spectra using an Airy function of FWHM of 12 {\AA} sampling every 5 {\AA} following the method described above. In the last block of Tables 2 to 5, we show the errors of the recovered H$\alpha$ and [{N\,\textsc{ii}}]$\lambda$6583 fluxes resulting from the convolutions.

Although the SDSS spectra also show the [{N\,\textsc{ii}}]$\lambda$6548 line in emission, since it is usually weak, it is not observed in the pseudo--spectra (see Fig. 9). To estimate the possible contamination of the [{N\,\textsc{ii}}]$\lambda$6548 line, we used the SDSS sample studied in \citet{Lara10a} for star forming galaxies up to z$\sim$0.1 (61921 galaxies), finding that the median flux of that line corresponds to the $\sim$10$\%$ of the median H$\alpha$ flux line of all the sample. Then, any contamination due to this line would be at most of the order of $\sim$2$\%$.

In the spectra of Fig. 9, we can observe that the flux errors are always lower than 20$\%$, which was the main goal of this study. Also, for the spSpec-53816-2231-307 galaxy it was possible to estimate their [{N\,\textsc{ii}}]$\lambda$6583 line flux with an error of $\sim 10\%$ although its flux is only 16$\%$ the {H$\alpha$} flux.


\begin{figure*}[h!]
\begin{minipage}[b]{1.1\linewidth}
\centering
\includegraphics[width=3cm,height=3cm]{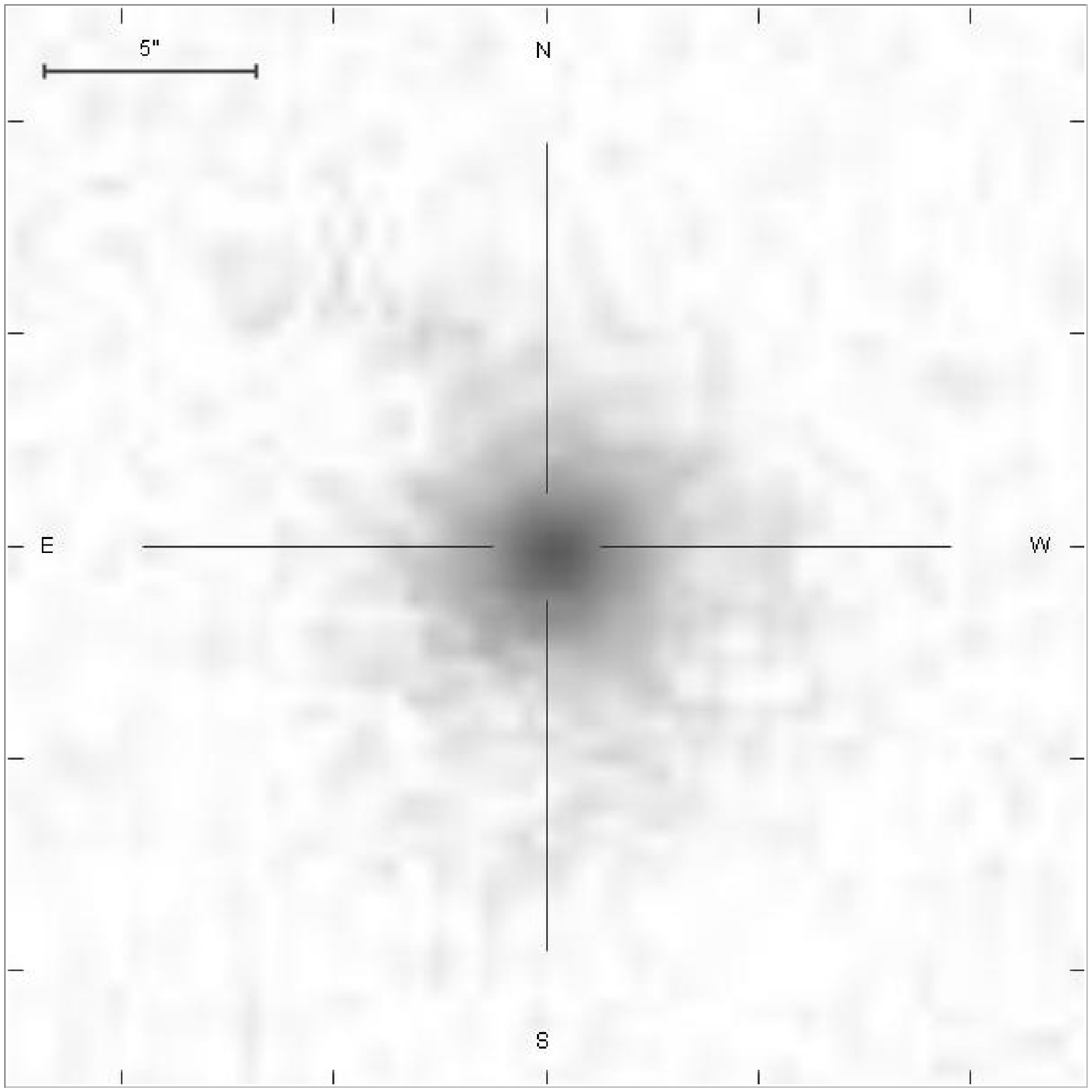}%
\hspace{1cm}
\includegraphics[width=3.7cm,height=3.3cm]{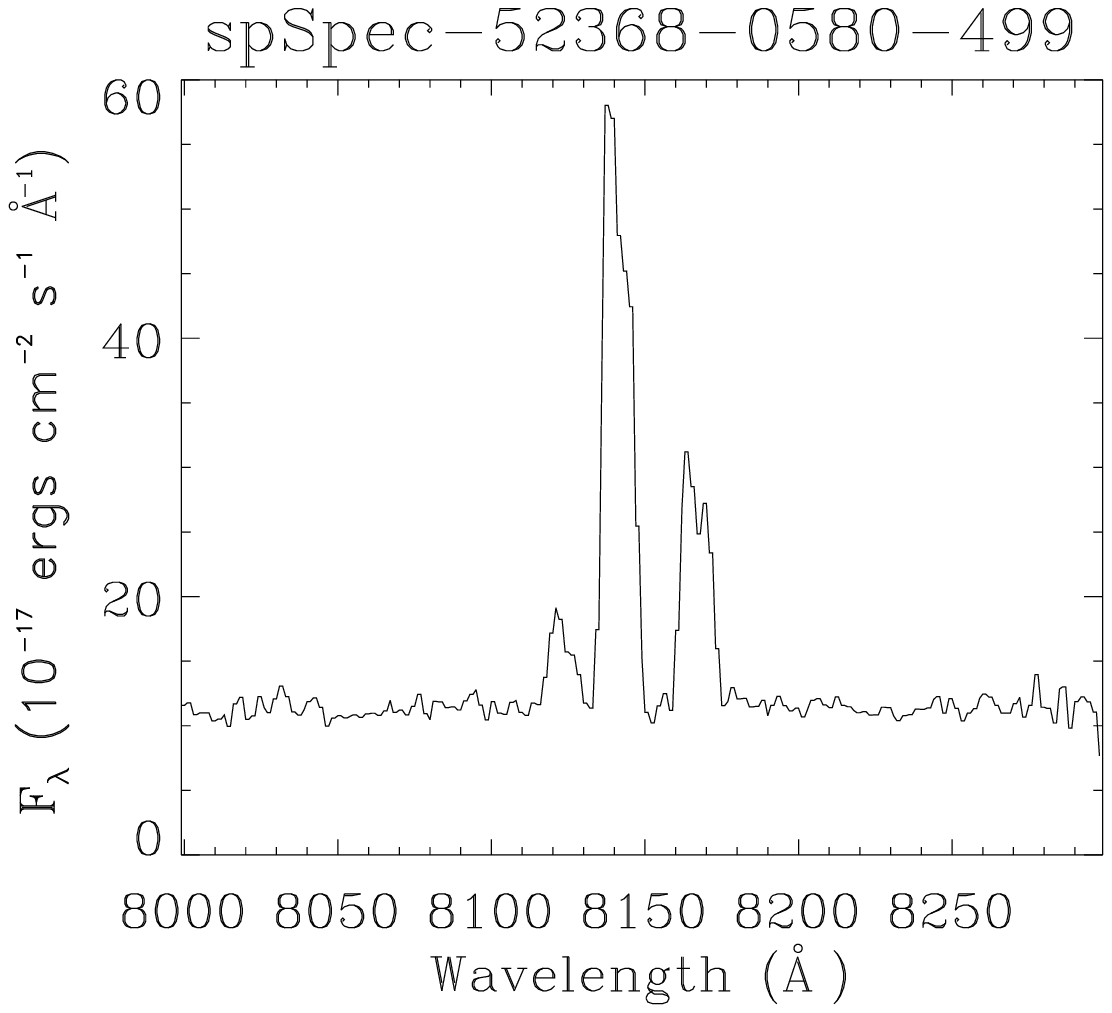}
\hspace{1cm}
\includegraphics[width=3.9cm,height=3.4cm]{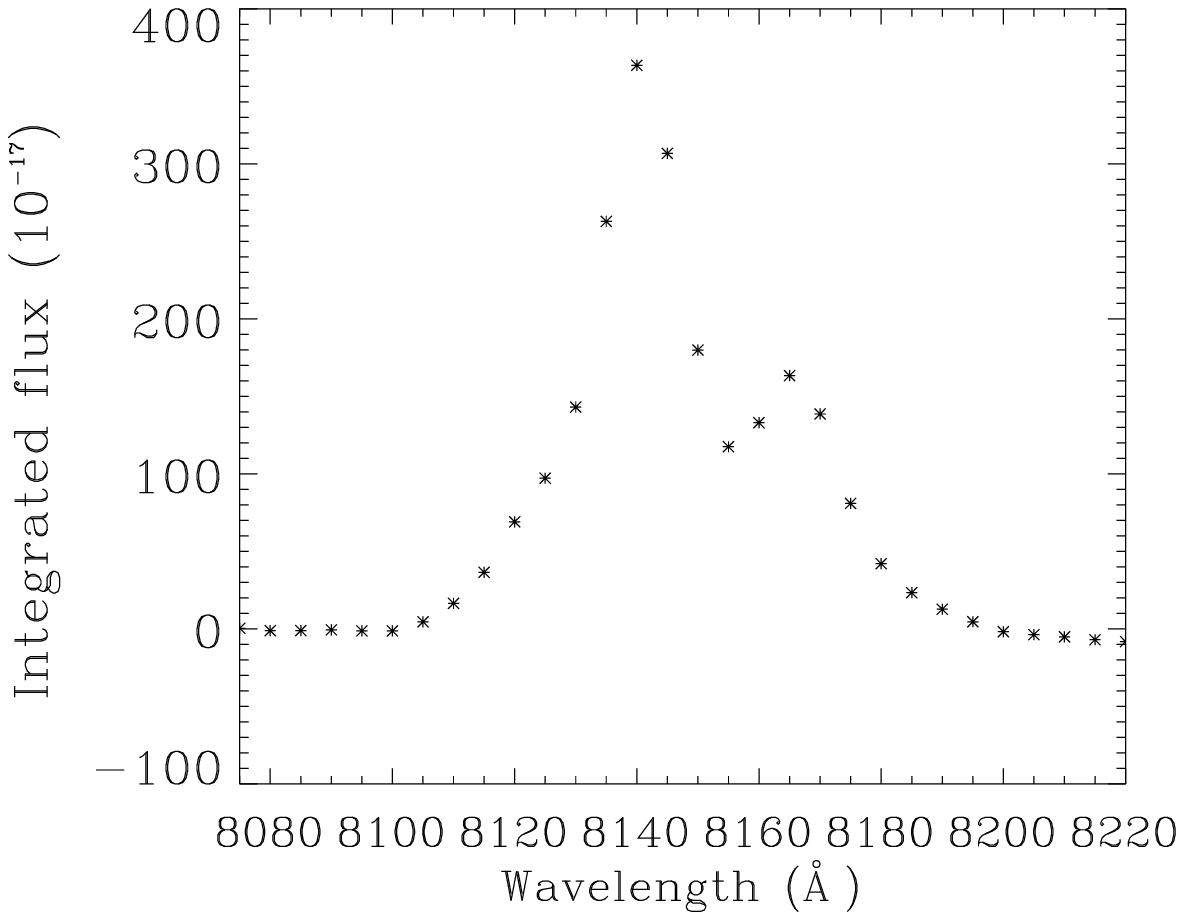}
\end{minipage}

\hspace{2cm}

\begin{minipage}[b]{1.1\linewidth}
\centering
\includegraphics[width=3cm,height=3cm]{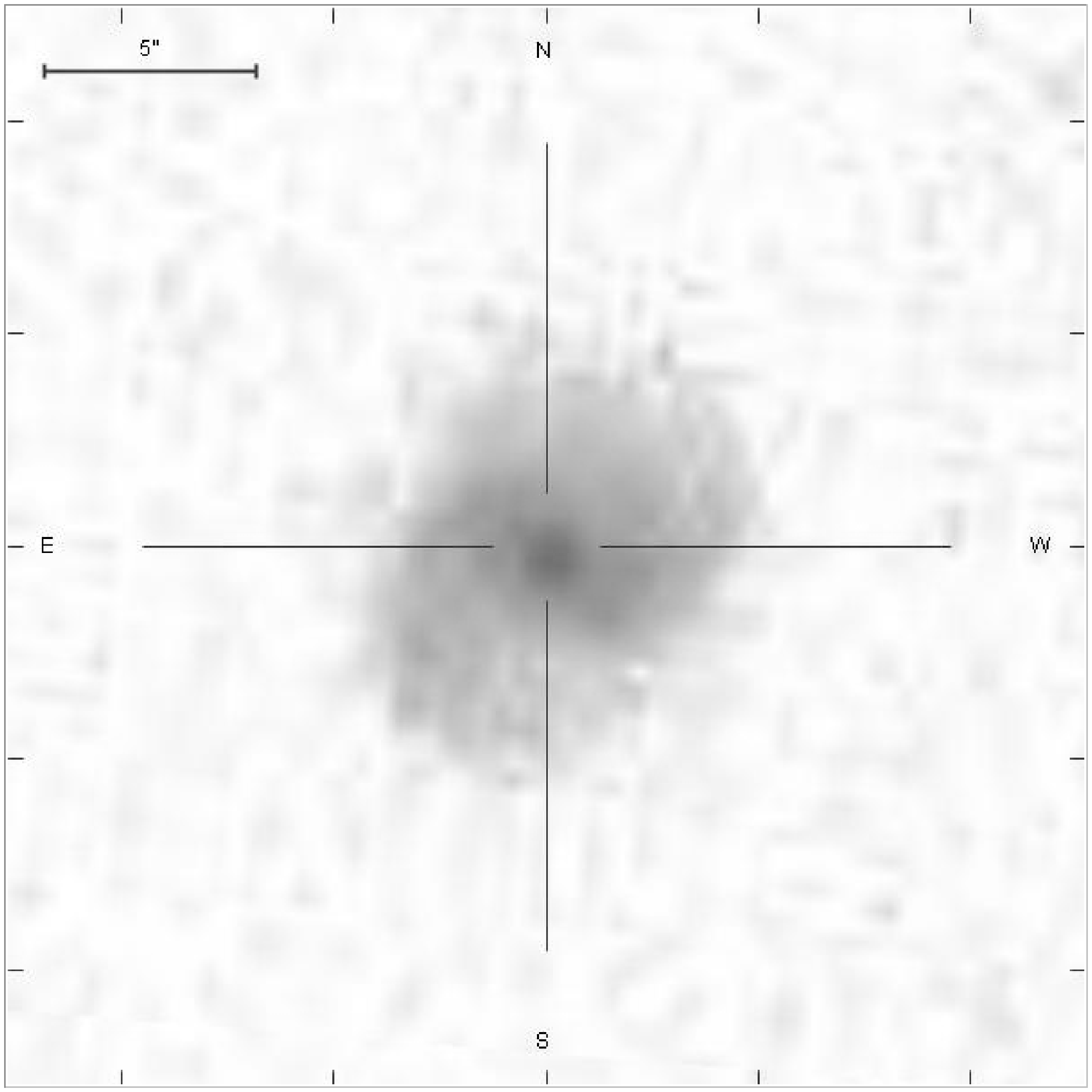}
\hspace{1cm}
\includegraphics[width=3.7cm,height=3.3cm]{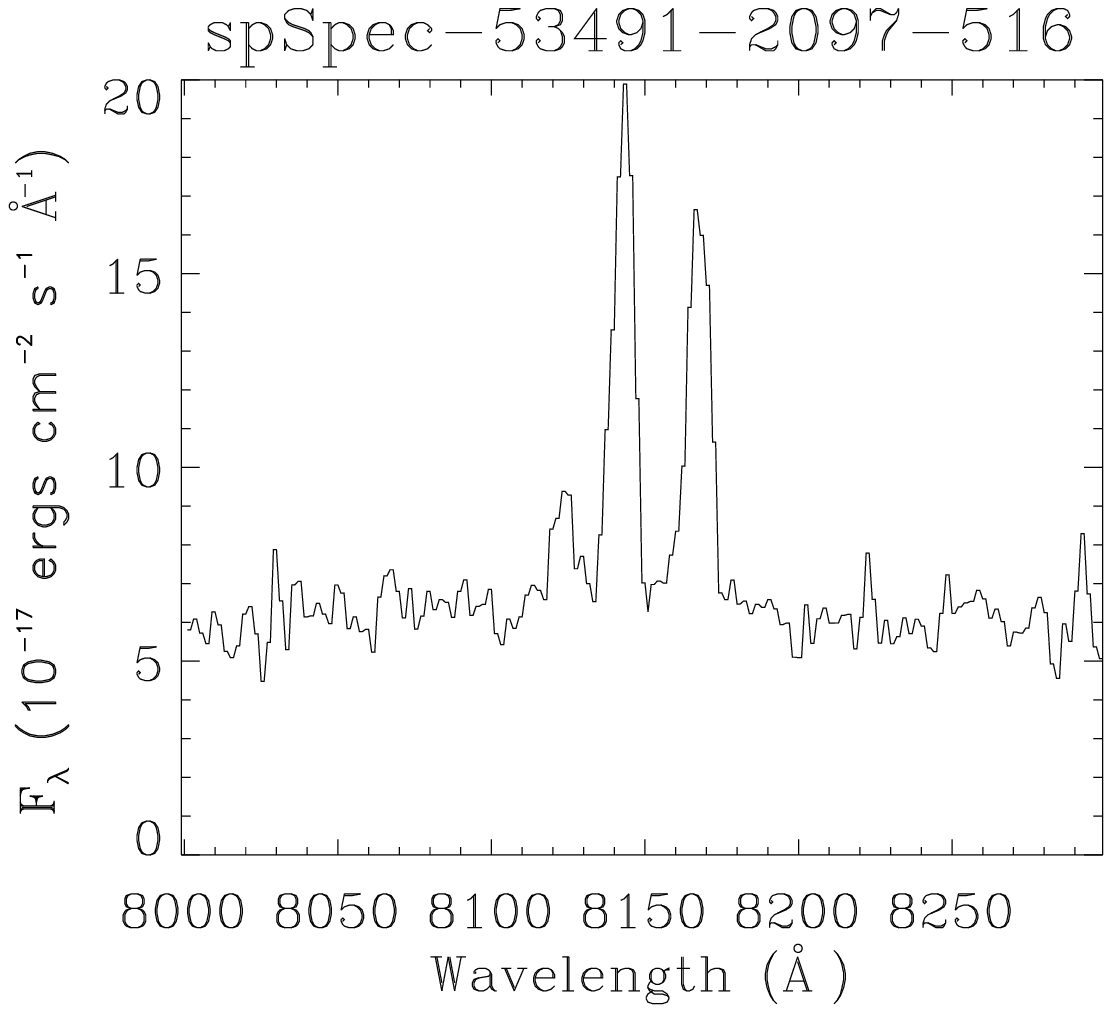}
\hspace{1cm}
\includegraphics[width=3.9cm,height=3.4cm]{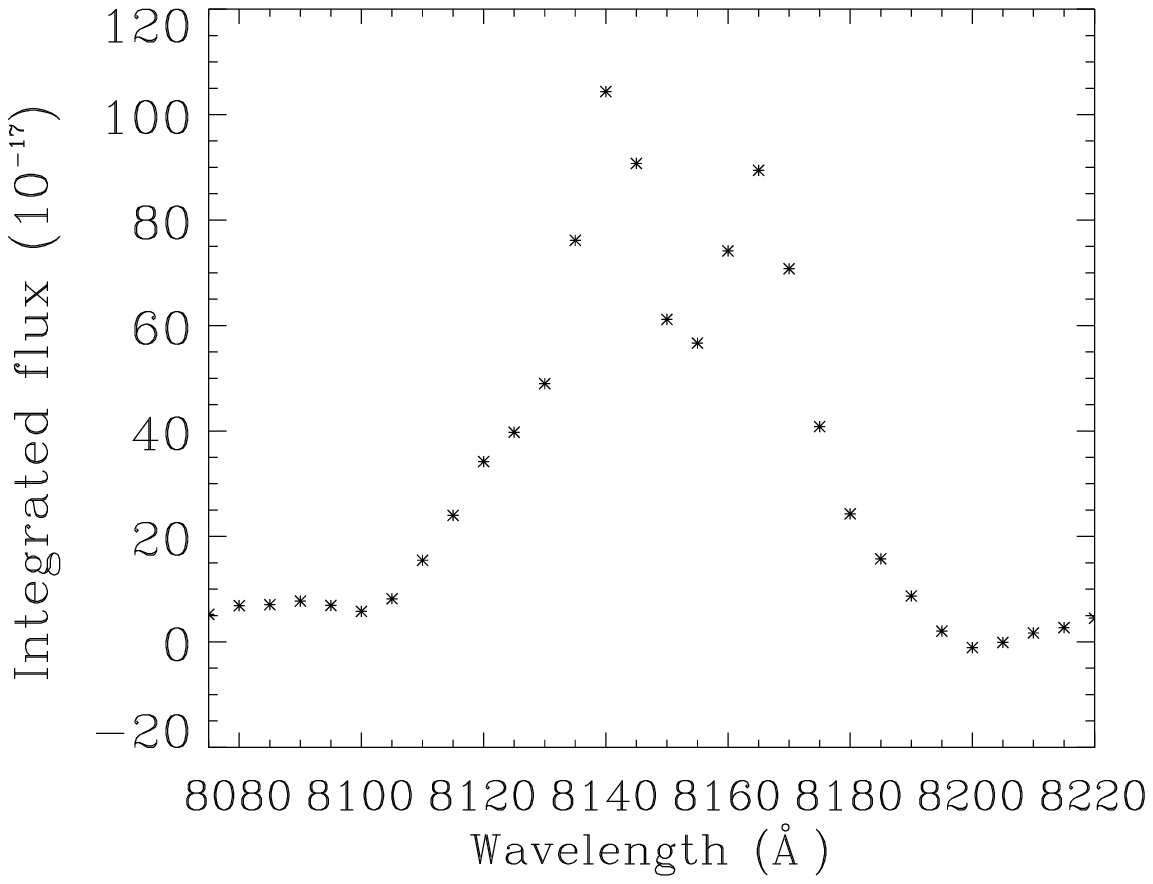}
\end{minipage}
\hspace{2cm}

\begin{minipage}[b]{1.1\linewidth}
\centering
\includegraphics[width=3cm,height=3cm]{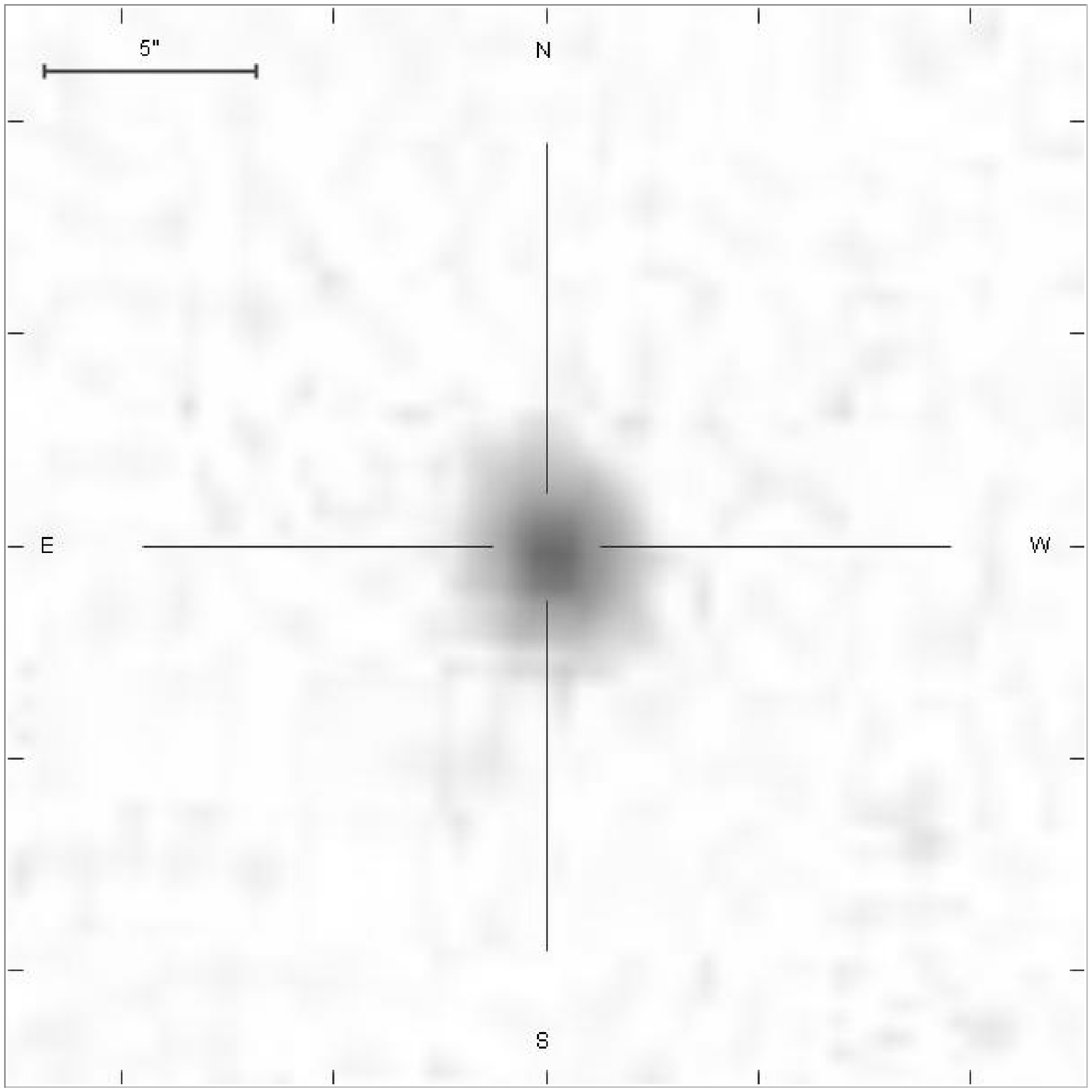}
\hspace{1cm}
\includegraphics[width=3.7cm,height=3.3cm]{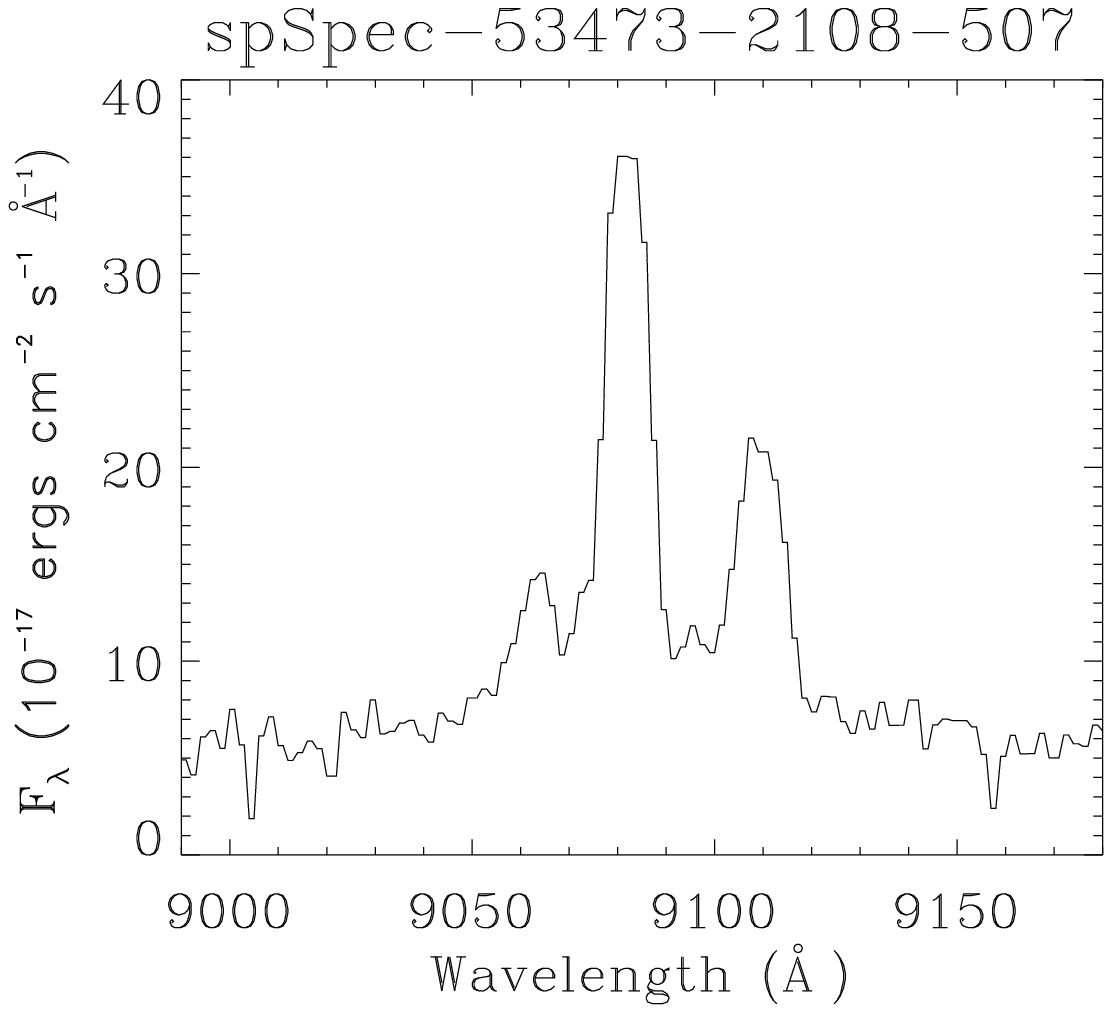}
\hspace{1cm}
\includegraphics[width=3.9cm,height=3.4cm]{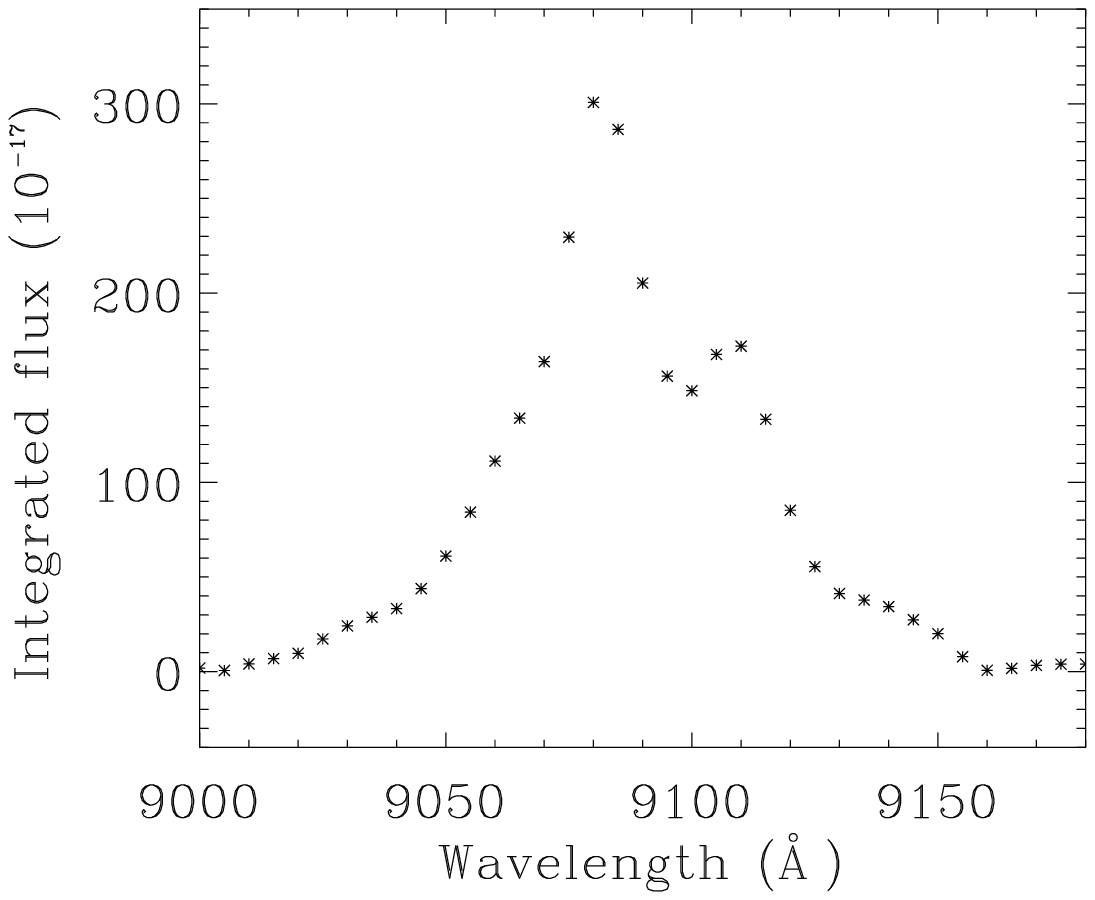}
\end{minipage}
\hspace{2cm}

\begin{minipage}[b]{1.1\linewidth}

\centering
\includegraphics[width=3cm,height=3cm]{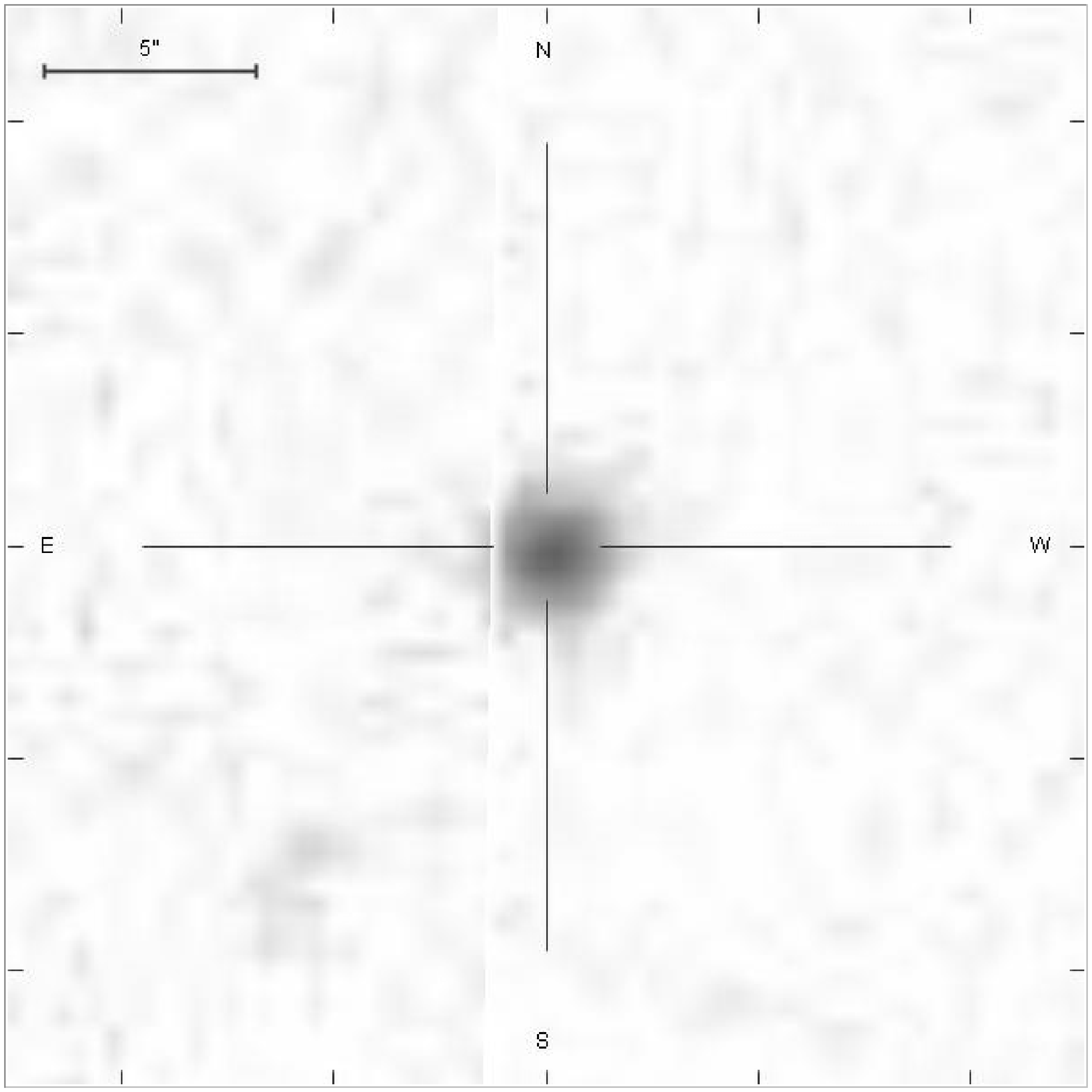}
\hspace{1cm}
\includegraphics[width=3.7cm,height=3.3cm]{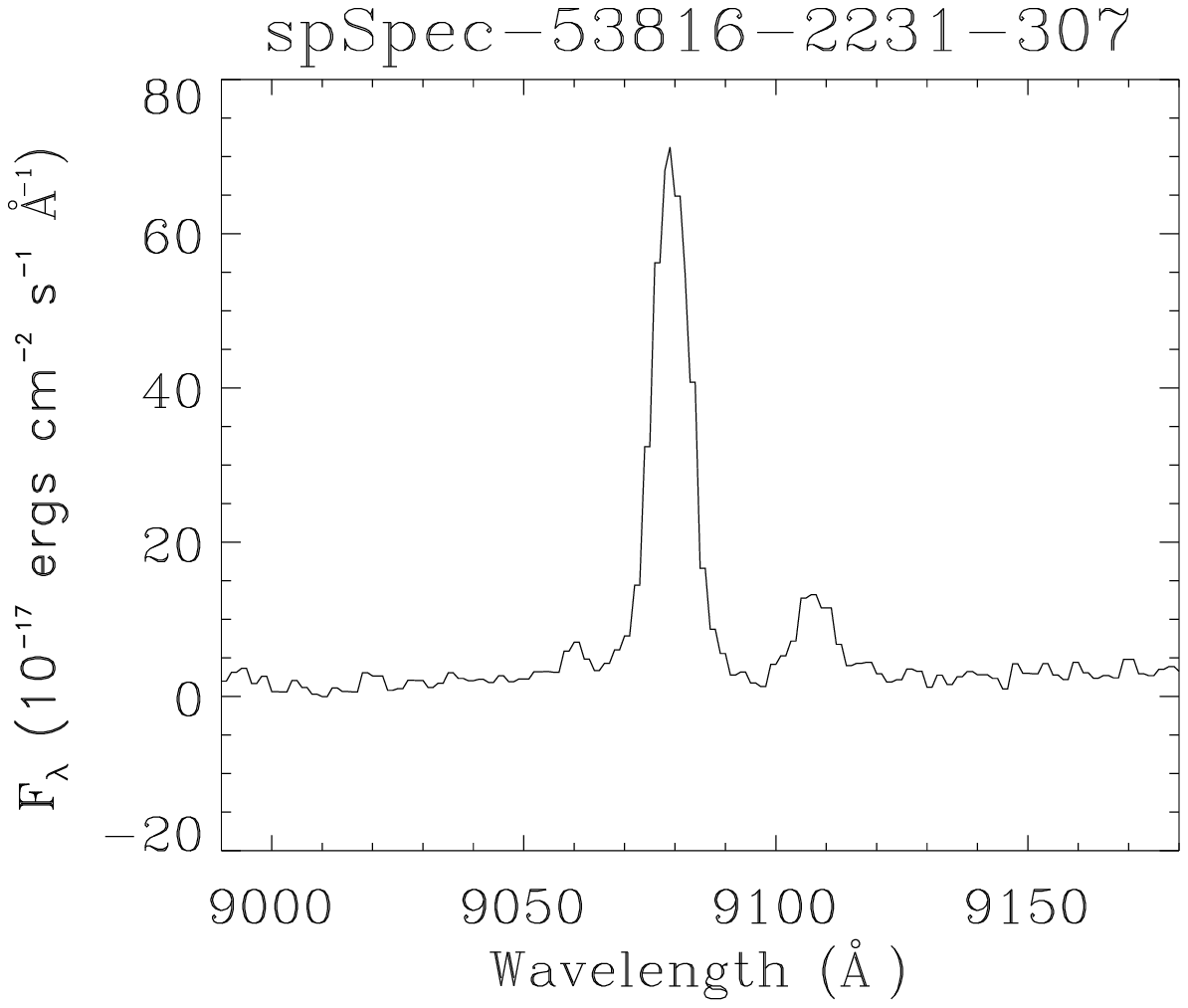}
\hspace{1cm}
\includegraphics[width=3.9cm,height=3.4cm]{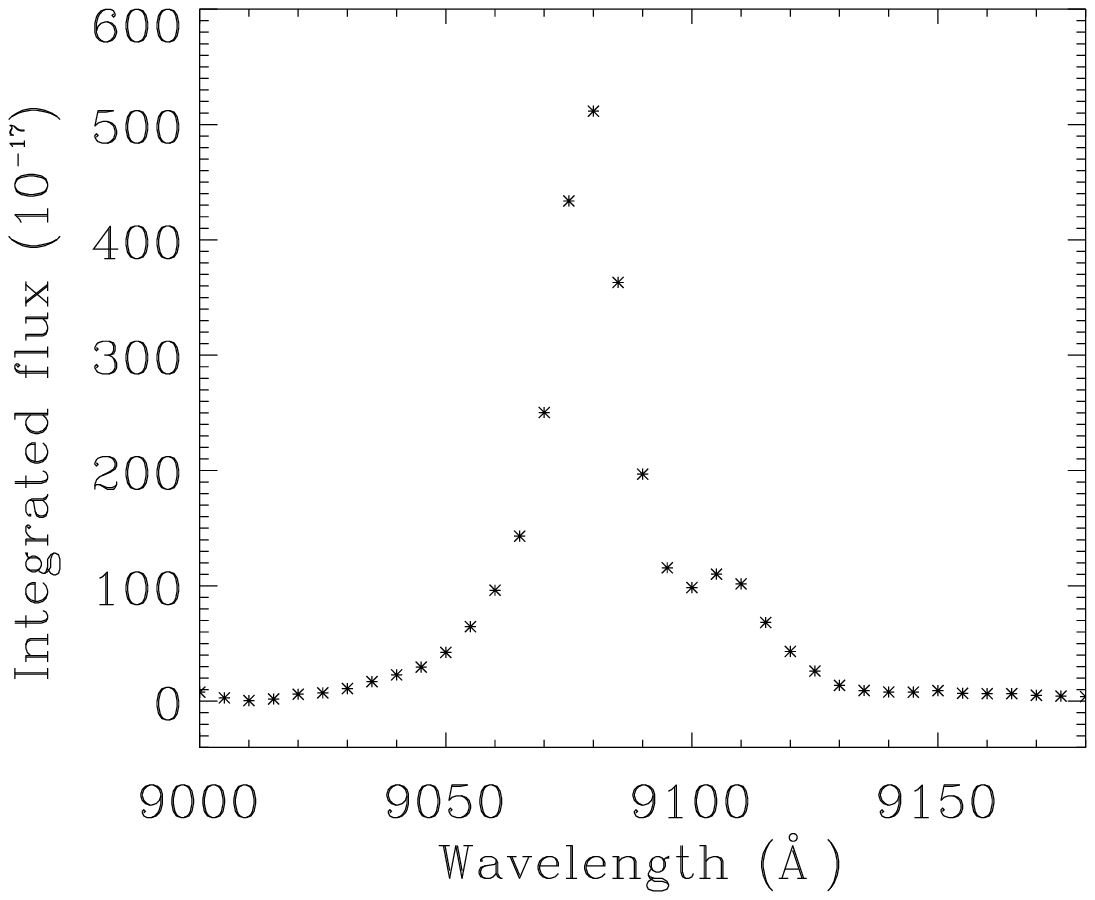}
\caption{From left to right: the images of the four galaxies selected from the SDSS to perform the convolutions (see text), the convolved section of their spectra showing H$\alpha$ and [{N\,\textsc{ii}}], and the result of the convolutions. Units as indicated in the text.}
\end{minipage}
\end{figure*}

\clearpage


\begin{table*}[h!]
\begin{center}
\small
\renewcommand{\arraystretch}{1.1}
\begin{tabular}{|c|c||c|c|c|c||c|c|c|c||c|c|c|}
\hline
\multicolumn{10}{|c||}{spSpec-52368-0580-499 (ra=164.85510, dec=4.77364)}&\multicolumn{3}{|c|}{Simulation result errors ($\%$)}\\\cline{1-13}
\multicolumn{2}{|c||}{}& \multicolumn{4}{|c||}{H$\alpha$}&\multicolumn{4}{|c||}{[{N\,\textsc{ii}}]}&\multicolumn{3}{|c|}{FWHM=12, sampling=5}\\\hline
$z$ &[{N\,\textsc{ii}}]/H$\alpha$ &Center & Height & $\sigma$ & Flux & Center & Height & $\sigma$ & Flux & H$\alpha$ error & [{N\,\textsc{ii}}] error &[{N\,\textsc{ii}}]/H$\alpha$ error \\\hline
0.2401 &0.44 & 8140.68 & 46.67 & 3.96 & 463.26 & 8166.34 & 18.93 & 4.3 & 204.04 & 17.50 & 12.20 & 6.80\\\hline
\end{tabular}
\end{center}
\caption{The table shows the galaxy redshift, [{N\,\textsc{ii}}]/H$\alpha$ ratio, and H$\alpha$ and [{N\,\textsc{ii}}] emision lines parameters for spSpec-52368-0580-499, as well as the flux error of these lines obtained from its convolution with a TF FWHM of 12{\AA}, sampling each 5{\AA}. Units as indicated in the text.}

 \end{table*}

 \begin{table*}[h!]
\begin{center}
\small
\renewcommand{\arraystretch}{1.1}
\begin{tabular}{|c|c||c|c|c|c||c|c|c|c||c|c|c|}
\hline
\multicolumn{10}{|c||}{spSpec-53491-2097-516 (ra=176.947073, dec=34.31164)}&\multicolumn{3}{|c|}{Simulation result errors ($\%$)}\\\cline{1-13}
\multicolumn{2}{|c||}{}& \multicolumn{4}{|c||}{H$\alpha$}&\multicolumn{4}{|c||}{[{N\,\textsc{ii}}]}&\multicolumn{3}{|c|}{FWHM=12, sampling=5}\\\hline
$z$& [{N\,\textsc{ii}}]/H$\alpha$ & Center & Height & $\sigma$ & Flux & Center & Height & $\sigma$ & Flux & H$\alpha$ error & [{N\,\textsc{ii}}] error &[{N\,\textsc{ii}}]/H$\alpha$ error \\\hline
0.2404&0.87 & 8142.48 & 12.85 & 3.39 & 109.19 & 8167.94 & 10.26 & 3.7 & 95.15 & 14.40 & 6.35 & 1.60\\\hline
\end{tabular}
\end{center}
\caption{The table shows the galaxy redshift, [{N\,\textsc{ii}}]/H$\alpha$ ratio, and H$\alpha$ and [{N\,\textsc{ii}}] emision lines parameters for spSpec-53491-2097-516, as well as the flux error of these lines obtained from its convolution with a TF FWHM of 12{\AA}, sampling each 5{\AA}. Units as indicated in the text.}
\end{table*}

\begin{table*}[h!]
\begin{center}
\small
\renewcommand{\arraystretch}{1.1}
\begin{tabular}{|c|c||c|c|c|c||c|c|c|c||c|c|c|}   
\hline
\multicolumn{10}{|c||}{spSpec-53473-2108-507 (ra=181.80184, dec=38.95954)}&\multicolumn{3}{|c|}{Simulation result errors ($\%$)}\\\cline{1-13}
\multicolumn{2}{|c||}{}& \multicolumn{4}{|c||}{H$\alpha$}&\multicolumn{4}{|c||}{[{N\,\textsc{ii}}]}&\multicolumn{3}{|c|}{FWHM=12, sampling=5}\\\hline
$z$& [{N\,\textsc{ii}}]/H$\alpha$ & Center & Height & $\sigma$ & Flux & Center & Height & $\sigma$ & Flux & H$\alpha$ error & [{N\,\textsc{ii}}] error &[{N\,\textsc{ii}}]/H$\alpha$ error \\\hline
0.3829 &0.61 & 9081.34 & 31.17 & 4.44 & 346.9 & 9108.04 & 14.44 & 5.9 &  213.55 & 13.30 & 19.60 & 7.20\\\hline
\end{tabular}
\end{center}
\caption{The table shows the galaxy redshift, [{N\,\textsc{ii}}]/H$\alpha$ ratio, and H$\alpha$ and [{N\,\textsc{ii}}] emision lines parameters for spSpec-53473-2108-507, as well as the flux error of these lines obtained from its convolution with a TF FWHM of 12{\AA}, sampling each 5{\AA}. Units as indicated in the text.}
\end{table*}

\begin{table*}[h!]
\begin{center}
\small
\renewcommand{\arraystretch}{1.1}
\begin{tabular}{|c|c||c|c|c|c||c|c|c|c||c|c|c|}   
\hline
\multicolumn{10}{|c||}{spSpec-53816-2231-307 (ra=183.29070, dec=27.09404)}&\multicolumn{3}{|c|}{Simulation result errors ($\%$)}\\\cline{1-13}
\multicolumn{2}{|c||}{}& \multicolumn{4}{|c||}{H$\alpha$}&\multicolumn{4}{|c||}{[{N\,\textsc{ii}}]}&\multicolumn{3}{|c|}{FWHM=12, sampling=5}\\\hline
$z$& [{N\,\textsc{ii}}]/H$\alpha$ & Center & Height & $\sigma$ & Flux & Center & Height & $\sigma$ & Flux & H$\alpha$ error & [{N\,\textsc{ii}}] error &[{N\,\textsc{ii}}]/H$\alpha$ error \\\hline
0.3830& 0.16 & 9078.72 & 70.91 & 3.32 & 590.1 & 9107.19 & 11.31 & 3.5 & 99.22 & 13.50 & 10.60 & 8.0\\\hline
\end{tabular}
\end{center}
\caption{The table shows the galaxy redshift, [{N\,\textsc{ii}}]/H$\alpha$ ratio, and H$\alpha$ and [{N\,\textsc{ii}}] emision lines parameters for spSpec-53816-2231-307, as well as the flux error of these lines obtained from its convolution with a TF FWHM of 12{\AA}, sampling each 5{\AA}. Units as indicated in the text.}
\end{table*}

\clearpage

\section{Summary and Conclusions}

In this work we generated spectra of typical star forming galaxies with different EWs (5, 10, 20, 30, 40, 50 {\AA}), at redshifts 0.24 and 0.4, which are the two windows of the OTELO survey for the {H$\alpha$} line. We convolved those spectra with the tunable filter response of the OSIRIS instrument of FWHM of 12 and 9{\AA}, subtracting the continuum, and estimating the relative errors of the recovered {H$\alpha$} and [{N\,\textsc{ii}}]$\lambda$6583 fluxes. We have concluded the following:

   \begin{itemize}

\item Using an Airy function with FWHM larger than 15 {\AA}, the errors of the recovered fluxes are larger than $\sim$ 25$\%$. Therefore, the convolutions were performed using a FWHM of 9 and 12{\AA}.

\item As a result of the convolutions, it was not possible to recover the FWHM of the {H$\alpha$} or [{N\,\textsc{ii}}]$\lambda$6583 lines, because the FWHM of the Airy function is larger than that of those lines. However, if the FWHM of any observed line is larger or of similar size than that of the convolved function (e.g., quasars), it will be possible to recover the FWHM of the observed line through a deconvolution. In those cases, to estimate the observed flux of the line, all the data--points of the pseudo--spectrum will be used.

\item The resulting pseudo--spectra show a decrement of the integrated flux at the edges, which is an effect of the limits of the wavelength window that in a real case would correspond to the wavelength limits of the order sorted used.

\item The initial wavelength and the sampling interval are of the highest importance, because both will determine how near the Airy function will be with respect to the observed emission line. The estimated flux error of the detected sources will be smaller when the peak of the Airy function is close to the peak of the emission line of the source.

\item The highest difference in wavelength of the pseudo--spectrum peak (detected line) with respect to the observed one, will be half of the sampling interval. This means that for a sampling of 6 {\AA}, the redshift error of the pseudo--spectra will be of $\Delta z=3$x$10^{-4}$. However, a fit to the pseudo--spectrum would reduce the error.

\item As a result of the convolutions, an Airy function with FWHM of 9 {\AA} allows minimizing contamination by closer lines, but generates large errors when recovering the {H$\alpha$} and [{N\,\textsc{ii}}]$\lambda$6583 fluxes. However, the error of the line ratio is smaller than that using a FWHM of 12{\AA}.

\item An Airy function with FWHM of 12 {\AA} produces smaller errors when recovering the fluxes of the lines. However, it favours the cross contamination of the fluxes of both lines. Also, the error of the line ratio {H$\alpha$}/[{N\,\textsc{ii}}]$\lambda$6583 is larger than that using a  FWHM of 9 {\AA}.


\item As a result of our simulations, we concluded that the combination of an Airy function of FWHM of 12 {\AA}, sampling every 5{\AA}, will allow separating the {H$\alpha$} and  [{N\,\textsc{ii}}]$\lambda$6583 emission lines with an error lower than 20$\%$. However, in the Appendix, the flux error estimates of the emission lines fluxes up to a sampling of 10 {\AA} are also presented. 

\item Although we selected the combination given above, according to Fig. 8, sampling every 5, 6, and 7 {\AA}, would give acceptable errors in the flux line measurements as well.

\item In order to test our method, we selected spectra from four SDSS-DR7 galaxies at redshifts 0.24 and 0.4, and convolved them with an Airy function of  FWHM of 12 {\AA} and sampling every 5{\AA}, obtaining in all cases errors lower than 20$\%$, even in those cases where the [{N\,\textsc{ii}}]$\lambda$6583 line was weak (e.g. [{N\,\textsc{ii}}]/{H$\alpha$}=0.16).

 \end{itemize}

As a result of our simulations we concluded that with the OSIRIS's TF is possible to estimate metallicities using the N2 method in galaxies spanning a wide range of EWs and morphological types, to discriminate star forming from AGN galaxies, and to estimate the SFR using the {H$\alpha$} flux. The selected combination of TF FWHM and sampling that will allow deblending {H$\alpha$} and  [{N\,\textsc{ii}}]$\lambda$6583 lines, and estimating their fluxes with an error lower than 20$\%$, is a TF FWHM of 12 {\AA} and a sampling of 5{\AA}.

\acknowledgments

This work was supported by the Spanish
\emph{Plan Nacional de Astronom\'{\i}a y Astrof\'{\i}sica} under grant AYA2008-06311-C02-01. The Sloan Digital Sky Survey (SDSS) is a joint project of The University of Chicago, Fermilab, the Institute for Advanced Study, the Japan Participation Group, The Johns Hopkins University, the Max--Planck--Institute for Astronomy, Princeton University, the United States Naval Observatory, and the University of Washington. Apache Point Observatory, site of the SDSS, is operated by the Astrophysical Research Consortium. Funding for the project has been provided by the Alfred P. Sloan Foundation, the SDSS member institutions, the National Aeronautics and Space Administration, the National Science Foundation, the U.S. Department of Energy, and Monbusho. The official SDSS web site is www.sdss.org. We thank the anonymous referee for all his/her constructive comments. Maritza A. Lara-L\'opez is supported by a CONACyT and SEP Mexican fellowships.


\clearpage

\onecolumn

\begin{table}[t!]
\begin{center}
\subtable{
\centering
\renewcommand{\arraystretch}{0.67}
\renewcommand\tabcolsep{7pt}
\begin{tabular}{c|c|c|c|c|c|c}   
\hline
\hline
\multicolumn{7}{c}{\bf FWHM of 12 {\AA}, $z$=0.24}\\\hline
EW  ({\AA})&    H$\alpha$ & $\sigma$ &[{N\,\textsc{ii}}]&  $\sigma$ &[{N\,\textsc{ii}}]/H$\alpha$&  $\sigma$\\\
{}&error ($\%$)& {} &error ($\%$)&  {} &error ($\%$)& {}\\\hline

\multicolumn{7}{c}{\bf Sampling: 3 {\AA}}\\\hline
50 & 9.06 & 3.94 & 3.95 & 3.35 & 12.34 & 5.05\\\hline
40 & 8.74 & 2.95 & 7.43 & 3.98 & 17.19 & 6.56\\\hline
30 & 9.30 & 2.79 & 5.38 & 5.08 & 11.88 & 8.21\\\hline
20 & 8.42 & 2.45 & 6.68 & 5.02 & 16.12 & 4.00\\\hline
10 & 9.26 & 3.11 & 7.05 & 6.96 & 17.23 & 6.06\\\hline
5 & 9.64 & 1.70 & 6.72 & 2.73 & 16.26 & 7.40\\\hline
 \multicolumn{7}{c}{\bf Sampling: 4 {\AA}}\\\hline
50 & 11.19 & 2.05 & 6.66 & 5.50 & 17.30 & 7.68\\\hline
40 & 10.87 & 3.05 & 8.43 & 5.57 & 19.25 & 10.28\\\hline
30 & 11.25 & 1.74 & 3.98 & 2.88 & 14.86 & 4.23\\\hline
20 & 9.06 & 3.67 & 4.47 & 3.17 & 8.12 & 6.65\\\hline
10 & 10.78 & 3.36 & 6.37 & 4.37 & 16.73 & 9.29\\\hline
5 & 9.28 & 2.27 & 4.51 & 3.46 & 12.31 & 6.81\\\hline
  \multicolumn{7}{c}{\bf Sampling: 5 {\AA}}\\\hline
50 & 11.47 & 3.49 & 6.93 & 4.55 & 19.29 & 9.79\\\hline
40 & 10.52 & 3.83 & 4.97 & 4.45 & 15.44 & 6.70\\\hline
30 & 10.68 & 2.48 & 6.36 & 4.92 & 16.64 & 8.53\\\hline
20 & 10.68 & 3.43 & 5.27 & 2.85 & 10.76 & 6.67\\\hline
10 & 13.06 & 5.15 & 3.41 & 2.31 & 14.50 & 7.54\\\hline
5 & 11.47 & 2.31 & 4.28 & 3.49 & 16.25 & 6.05\\\hline
  \multicolumn{7}{c}{\bf Sampling: 6 {\AA}}\\\hline
50 & 10.52 & 4.53 & 4.96 & 3.12 & 15.33 & 9.55\\\hline
40 & 13.35 & 5.11 & 5.55 & 3.37 & 18.55 & 9.36\\\hline
30 & 13.54 & 5.78 & 4.17 & 3.25 & 16.51 & 7.18\\\hline
20 & 14.33 & 4.12 & 4.98 & 3.05 & 18.55 & 7.49\\\hline
10 & 12.71 & 4.13 & 5.90 & 5.44 & 16.26 & 11.37\\\hline
5 & 11.28 & 4.96 & 5.34 & 3.24 & 10.13 & 8.31\\\hline
  \multicolumn{7}{c}{\bf Sampling: 7 {\AA}}\\\hline
50 & 11.37 & 4.99 & 6.05 & 4.42 & 9.75 & 8.47\\\hline
40 & 12.07 & 6.35 & 6.02 & 4.33 & 17.34 & 12.12\\\hline
30 & 14.23 & 5.95 & 6.97 & 3.92 & 19.46 & 14.41\\\hline
20 & 12.00 & 5.17 & 8.04 & 2.89 & 13.73 & 12.44\\\hline
10 & 13.65 & 6.40 & 6.34 & 4.26 & 14.86 & 13.74\\\hline
5 & 12.32 & 4.08 & 6.56 & 3.81 & 16.04 & 12.41\\\hline
  \multicolumn{7}{c}{\bf Sampling: 8 {\AA}}\\\hline
50 & 16.28 & 7.57 & 4.17 & 2.59 & 17.35 & 7.39\\\hline
40 & 15.32 & 5.60 & 7.69 & 6.07 & 16.35 & 10.54\\\hline
30 & 14.52 & 8.10 & 7.61 & 6.76 & 14.72 & 10.32\\\hline
20 & 15.32 & 7.11 & 5.43 & 4.13 & 17.67 & 9.26\\\hline
10 & 15.00 & 5.58 & 7.79 & 4.08 & 13.51 & 8.99\\\hline
5 & 13.41 & 5.10 & 7.07 & 5.13 & 11.27 & 6.86\\\hline
  \multicolumn{7}{c}{\bf Sampling: 9 {\AA}}\\\hline
50 & 17.29 & 8.08 & 7.94 & 6.75 & 16.17 & 14.70\\\hline
40 & 14.91 & 7.99 & 7.77 & 5.11 & 14.52 & 12.81\\\hline
30 & 15.07 & 8.99 & 8.56 & 4.76 & 16.13 & 11.31\\\hline
20 & 17.13 & 7.70 & 5.18 & 3.84 & 18.43 & 13.79\\\hline
10 & 13.80 & 6.64 & 7.36 & 6.50 & 14.73 & 9.44\\\hline
5 & 16.49 & 6.26 & 5.33 & 5.00 & 18.25 & 13.21\\\hline
  \multicolumn{7}{c}{\bf Sampling: 10 {\AA}}\\\hline
50 & 18.36 & 7.77 & 7.62 & 7.98 & 24.71 & 16.23\\\hline
40 & 17.44 & 7.91 & 8.11 & 6.06 & 20.37 & 19.00\\\hline
30 & 17.12 & 9.09 & 7.90 & 5.79 & 18.53 & 17.83\\\hline
20 & 19.47 & 8.67 & 8.78 & 6.72 & 22.99 & 17.71\\\hline
10 & 18.23 & 9.84 & 9.37 & 7.81 & 21.80 & 18.74\\\hline
5 & 17.44 & 10.33 & 9.81 & 7.55 & 25.88 & 20.19\\\hline

\end{tabular}}
\caption{Average H$\alpha$, [{N\,\textsc{ii}}], and [{N\,\textsc{ii}}]/H$\alpha$ error with its respective sigma. Errors were estimated for a tunable filter FWHM of 12 {\AA} using the simulated spectra at redshift 0.24}

\end{center}
\end{table}

\clearpage

\begin{table}[t!]
\begin{center}
\subtable{
\centering
\renewcommand{\arraystretch}{0.67}
\renewcommand\tabcolsep{7pt}
\begin{tabular}{c|c|c|c|c|c|c}   
\hline
\hline

\multicolumn{7}{c}{\bf FWHM of 12 {\AA}, $z$=0.4}\\\hline
EW  ({\AA})&    H$\alpha$ & $\sigma$ & [{N\,\textsc{ii}}] &  $\sigma$ &  [{N\,\textsc{ii}}]/H$\alpha$ &  $\sigma$\\\
{}&error ($\%$)& {} &error ($\%$)&  {} &error ($\%$)& {}\\\hline

\multicolumn{7}{c}{\bf Sampling: 3 {\AA}}\\\hline
50 & 11.81 & 1.79 & 4.66 & 2.93 & 11.59 & 5.62\\\hline
40 & 10.92 & 2.48 & 4.91 & 3.78 & 10.57 & 4.89\\\hline
30 & 11.71 & 1.78 & 3.28 & 2.08 & 12.84 & 5.41\\\hline
20 & 10.85 & 3.09 & 4.32 & 3.76 & 10.95 & 8.40\\\hline
10 & 12.07 & 2.08 & 4.52 & 3.74 & 12.62 & 6.97\\\hline
5 & 11.23 & 2.82 & 5.72 & 3.56 & 10.84 & 5.78\\\hline
\multicolumn{7}{c}{\bf Sampling: 4 {\AA}}\\\hline
50 & 13.60 & 1.25 & 4.59 & 3.60 & 16.20 & 5.70\\\hline
40 & 11.17 & 2.00 & 3.52 & 2.53 & 12.60 & 4.73\\\hline
30 & 12.34 & 1.73 & 4.96 & 2.84 & 12.82 & 5.61\\\hline
20 & 13.64 & 1.90 & 4.20 & 2.85 & 12.32 & 4.90\\\hline
10 & 13.83 & 2.47 & 4.77 & 2.83 & 16.16 & 7.04\\\hline
5 & 12.47 & 2.86 & 4.09 & 3.27 & 11.98 & 5.75\\\hline
\multicolumn{7}{c}{\bf Sampling: 5 {\AA}}\\\hline
50 & 14.32 & 3.41 & 8.01 & 4.55 & 10.25 & 10.29\\\hline
40 & 12.92 & 3.64 & 6.01 & 3.14 & 12.46 & 7.49\\\hline
30 & 13.20 & 4.88 & 4.92 & 5.05 & 12.65 & 9.17\\\hline
20 & 14.40 & 2.93 & 6.26 & 3.99 & 13.82 & 10.60\\\hline
10 & 13.34 & 4.54 & 6.44 & 4.34 & 13.48 & 10.49\\\hline
5 & 14.96 & 3.24 & 5.17 & 4.57 & 13.94 & 7.73\\\hline
\multicolumn{7}{c}{\bf Sampling: 6 {\AA}}\\\hline
50 & 15.81 & 3.87 & 4.73 & 4.03 & 14.75 & 9.49\\\hline
40 & 12.73 & 4.16 & 5.02 & 4.61 & 10.94 & 8.21\\\hline
30 & 13.52 & 3.62 & 8.44 & 6.09 & 14.16 & 13.03\\\hline
20 & 16.28 & 4.32 & 10.18 & 3.96 & 12.08 & 14.18\\\hline
10 & 15.53 & 3.65 & 4.75 & 3.91 & 14.67 & 8.16\\\hline
5 & 14.22 & 3.66 & 8.95 & 5.69 & 13.47 & 9.56\\\hline
\multicolumn{7}{c}{\bf Sampling: 7 {\AA}}\\\hline
50 & 15.69 & 4.99 & 5.60 & 5.32 & 12.60 & 3.10\\\hline
40 & 15.69 & 3.35 & 5.82 & 4.57 & 15.74 & 9.04\\\hline
30 & 14.57 & 4.24 & 6.66 & 3.90 & 13.27 & 8.94\\\hline
20 & 14.93 & 5.86 & 7.86 & 6.15 & 14.53 & 11.44\\\hline
10 & 15.27 & 5.93 & 6.29 & 5.09 & 13.02 & 8.29\\\hline
5 & 13.73 & 4.55 & 6.49 & 5.46 & 11.55 & 7.67\\\hline
\multicolumn{7}{c}{\bf Sampling: 8 {\AA}}\\\hline
50 & 18.84 & 5.91 & 6.00 & 4.49 & 18.46 & 11.70\\\hline
40 & 15.61 & 6.20 & 9.11 & 6.30 & 14.77 & 8.01\\\hline
30 & 17.01 & 5.24 & 8.99 & 5.82 & 12.31 & 9.64\\\hline
20 & 18.00 & 7.27 & 9.95 & 5.11 & 15.28 & 10.99\\\hline
10 & 17.43 & 6.48 & 9.63 & 5.95 & 14.05 & 9.43\\\hline
5 & 16.63 & 5.28 & 9.87 & 6.39 & 15.39 & 10.66\\\hline
\multicolumn{7}{c}{\bf Sampling: 9 {\AA}}\\\hline
50 & 18.94 & 5.94 & 11.58 & 6.48 & 13.88 & 13.09\\\hline
40 & 17.68 & 7.14 & 10.49 & 4.78 & 15.17 & 9.55\\\hline
30 & 20.34 & 6.72 & 7.76 & 5.73 & 17.02 & 8.81\\\hline
20 & 18.80 & 7.15 & 7.12 & 5.40 & 16.42 & 9.47\\\hline
10 & 17.68 & 4.10 & 6.68 & 5.50 & 16.26 & 6.30\\\hline
5 & 19.78 & 7.73 & 8.20 & 6.53 & 18.72 & 11.02\\\hline
\multicolumn{7}{c}{\bf Sampling: 10 {\AA}}\\\hline
50 & 20.32 & 8.53 & 12.28 & 8.72 & 20.53 & 17.39\\\hline
40 & 21.30 & 6.97 & 13.27 & 8.53 & 12.39 & 8.26\\\hline
30 & 19.76 & 9.48 & 8.97 & 6.11 & 15.27 & 15.17\\\hline
20 & 22.14 & 7.96 & 11.99 & 9.79 & 22.34 & 16.67\\\hline
10 & 19.91 & 9.42 & 9.96 & 7.21 & 18.09 & 16.27\\\hline
5 & 22.01 & 7.47 & 10.13 & 7.76 & 21.02 & 16.27\\\hline

\end{tabular} }
\caption{Average H$\alpha$, [{N\,\textsc{ii}}], and [{N\,\textsc{ii}}]/H$\alpha$ error with its respective sigma. Errors were estimated for a tunable filter FWHM of 12 using the simulated spectra at redshift 0.4}
\end{center}
\end{table}

\clearpage
\begin{table}[t!]
\begin{center}
\subtable{
\centering
\renewcommand{\arraystretch}{0.67}
\renewcommand\tabcolsep{7pt}
\begin{tabular}{c|c|c|c|c|c|c}   
\hline
\hline
\multicolumn{7}{c}{\bf FWHM of 9 {\AA}, $z$=0.24}\\\hline
EW  ({\AA})&    H$\alpha$ & $\sigma$ &[{N\,\textsc{ii}}]&  $\sigma$ &[{N\,\textsc{ii}}]/H$\alpha$&  $\sigma$\\\
{}&error ($\%$)& {} &error ($\%$)&  {} &error ($\%$)& {}\\\hline

\multicolumn{7}{c}{\bf Sampling: 3 {\AA}}\\\hline
50 & 14.17 & 2.80 & 8.74 & 5.99 & 9.52 & 5.88\\\hline
40 & 16.38 & 2.99 & 6.28 & 3.80 & 15.33 & 8.80\\\hline
30 & 15.60 & 4.53 & 9.19 & 4.06 & 9.05 & 6.13\\\hline
20 & 15.92 & 3.56 & 11.11 & 4.57 & 6.85 & 8.35\\\hline
10 & 15.28 & 3.83 & 12.04 & 5.41 & 10.15 & 9.65\\\hline
5 & 16.32 & 3.11 & 11.45 & 4.49 & 8.89 & 6.00\\\hline
\multicolumn{7}{c}{\bf Sampling: 4 {\AA}}\\\hline
50 & 18.23 & 2.66 & 10.50 & 4.76 & 9.62 & 7.75\\\hline
40 & 18.16 & 4.71 & 10.51 & 5.65 & 9.55 & 6.58\\\hline
30 & 18.95 & 3.25 & 11.16 & 6.42 & 11.45 & 9.13\\\hline
20 & 19.19 & 2.99 & 10.50 & 5.14 & 10.87 & 7.30\\\hline
10 & 17.60 & 2.85 & 9.96 & 4.62 & 9.42 & 6.61\\\hline
5 & 17.04 & 3.70 & 11.21 & 5.41 & 8.89 & 6.61\\\hline
\multicolumn{7}{c}{\bf Sampling: 5 {\AA}}\\\hline
50 & 18.38 & 4.87 & 12.22 & 6.23 & 9.07 & 8.78\\\hline
40 & 18.85 & 4.71 & 12.65 & 6.71 & 8.79 & 7.91\\\hline
30 & 18.85 & 3.75 & 13.49 & 5.24 & 9.34 & 12.20\\\hline
20 & 19.33 & 4.09 & 11.25 & 5.65 & 10.08 & 5.79\\\hline
10 & 17.63 & 5.75 & 10.91 & 7.50 & 10.77 & 4.60\\\hline
5 & 18.69 & 4.82 & 13.74 & 4.98 & 7.18 & 4.35\\\hline
\multicolumn{7}{c}{\bf Sampling: 6 {\AA}}\\\hline
50 & 20.76 & 5.45 & 13.27 & 8.23 & 11.00 & 7.33\\\hline
40 & 20.69 & 6.10 & 16.40 & 8.34 & 7.14 & 5.99\\\hline
30 & 22.03 & 5.74 & 18.74 & 7.98 & 10.98 & 7.00\\\hline
20 & 20.05 & 5.23 & 12.52 & 8.46 & 11.26 & 9.25\\\hline
10 & 22.27 & 5.40 & 15.47 & 8.31 & 13.99 & 8.45\\\hline
5 & 21.23 & 4.93 & 15.17 & 7.01 & 9.25 & 6.66\\\hline
\multicolumn{7}{c}{\bf Sampling: 7 {\AA}}\\\hline
50 & 20.33 & 7.72 & 16.89 & 6.65 & 8.96 & 12.78\\\hline
40 & 22.40 & 4.82 & 15.36 & 7.03 & 12.58 & 11.40\\\hline
30 & 21.44 & 6.49 & 15.44 & 7.98 & 15.94 & 11.02\\\hline
20 & 20.49 & 4.52 & 15.38 & 7.87 & 13.86 & 11.92\\\hline
10 & 20.65 & 5.89 & 12.59 & 7.35 & 16.99 & 14.15\\\hline
5 & 20.03 & 4.97 & 18.70 & 9.09 & 12.74 & 10.60\\\hline
\multicolumn{7}{c}{\bf Sampling: 8 {\AA}}\\\hline
50 & 22.66 & 8.63 & 14.29 & 7.13 & 13.22 & 12.35\\\hline
40 & 24.16 & 5.18 & 14.95 & 7.48 & 12.14 & 5.90\\\hline
30 & 24.00 & 8.66 & 13.93 & 6.62 & 14.09 & 11.20\\\hline
20 & 23.05 & 8.47 & 15.74 & 8.13 & 12.04 & 11.86\\\hline
10 & 24.48 & 6.98 & 13.98 & 9.45 & 17.54 & 9.85\\\hline
5 & 21.62 & 9.75 & 15.67 & 10.96 & 13.97 & 6.84\\\hline
\multicolumn{7}{c}{\bf Sampling: 9 {\AA}}\\\hline
50 & 25.85 & 10.02 & 16.57 & 9.97 & 22.94 & 22.84\\\hline
40 & 25.69 & 9.47 & 16.77 & 12.07 & 21.22 & 23.18\\\hline
30 & 24.10 & 9.79 & 18.41 & 9.45 & 17.99 & 13.33\\\hline
20 & 25.69 & 8.98 & 18.77 & 10.54 & 16.15 & 12.10\\\hline
10 & 24.74 & 7.92 & 20.21 & 10.20 & 14.94 & 8.03\\\hline
5 & 24.74 & 10.23 & 16.60 & 9.06 & 19.21 & 16.75\\\hline
\multicolumn{7}{c}{\bf Sampling: 10 {\AA}}\\\hline
50 & 27.66 & 11.46 & 21.26 & 9.76 & 25.85 & 19.39\\\hline
40 & 27.55 & 11.64 & 25.60 & 13.67 & 29.98 & 20.31\\\hline
30 & 25.80 & 11.30 & 23.07 & 13.54 & 28.85 & 20.31\\\hline
20 & 27.50 & 11.46 & 20.90 & 9.22 & 24.64 & 18.40\\\hline
10 & 27.34 & 10.98 & 20.33 & 11.78 & 28.46 & 19.98\\\hline
5 & 28.77 & 10.03 & 17.87 & 11.15 & 32.03 & 25.38\\\hline

\end{tabular} }
\caption{Average H$\alpha$, [{N\,\textsc{ii}}], and [{N\,\textsc{ii}}]/H$\alpha$ error with its respective sigma. Errors were estimated for a tunable filter FWHM of 9{\AA} using the simulated spectra at redshift 0.24}
\end{center}
\end{table}

\clearpage
\begin{table}[t!]
\begin{center}
\subtable{
\centering
\renewcommand{\arraystretch}{0.67}
\renewcommand\tabcolsep{7pt}
\begin{tabular}{c|c|c|c|c|c|c}   
\hline
\hline
\multicolumn{7}{c}{\bf FWHM of 9 {\AA}, $z$=0.4}\\\hline
EW  ({\AA})&    H$\alpha$ & $\sigma$ & [{N\,\textsc{ii}}] &  $\sigma$ &  [{N\,\textsc{ii}}]/H$\alpha$ &  $\sigma$\\\
{}&error ($\%$)& {} &error ($\%$)&  {} &error ($\%$)& {}\\\hline

\multicolumn{7}{c}{\bf Sampling: 3 {\AA}}\\\hline
50 & 19.90 & 2.64 & 13.43 & 5.97 & 8.29 & 7.21\\\hline
40 & 19.90 & 2.45 & 14.77 & 4.39 & 6.84 & 3.95\\\hline
30 & 20.34 & 2.23 & 14.50 & 5.39 & 7.44 & 6.32\\\hline
20 & 19.08 & 3.13 & 13.82 & 5.32 & 8.52 & 4.92\\\hline
10 & 20.46 & 2.53 & 14.98 & 3.11 & 8.28 & 4.62\\\hline
5 & 20.18 & 2.47 & 10.57 & 4.95 & 12.06 & 5.87\\\hline
\multicolumn{7}{c}{\bf Sampling: 4 {\AA}}\\\hline
50 & 20.70 & 2.97 & 16.85 & 5.62 & 6.50 & 6.16\\\hline
40 & 20.77 & 2.79 & 17.61 & 2.93 & 6.10 & 3.03\\\hline
30 & 20.35 & 2.94 & 15.42 & 4.59 & 6.69 & 5.82\\\hline
20 & 22.10 & 2.29 & 13.23 & 4.80 & 12.08 & 6.28\\\hline
10 & 21.05 & 3.17 & 11.33 & 6.33 & 12.27 & 5.80\\\hline
5 & 20.98 & 3.84 & 13.84 & 5.78 & 10.56 & 5.12\\\hline
\multicolumn{7}{c}{\bf Sampling: 5 {\AA}}\\\hline
50 & 22.49 & 2.27 & 13.21 & 5.48 & 12.31 & 8.56\\\hline
40 & 21.54 & 3.31 & 13.85 & 5.93 & 10.44 & 7.40\\\hline
30 & 22.05 & 3.59 & 16.39 & 6.03 & 10.88 & 9.19\\\hline
20 & 23.17 & 2.94 & 18.95 & 4.45 & 8.14 & 4.75\\\hline
10 & 22.47 & 2.74 & 12.90 & 4.92 & 12.53 & 8.56\\\hline
5 & 22.89 & 1.54 & 19.58 & 5.42 & 8.16 & 3.32\\\hline
\multicolumn{7}{c}{\bf Sampling: 6 {\AA}}\\\hline
50 & 23.90 & 3.73 & 17.34 & 5.32 & 10.90 & 8.02\\\hline
40 & 23.90 & 5.93 & 16.39 & 8.79 & 14.54 & 11.84\\\hline
30 & 24.18 & 5.47 & 16.47 & 5.07 & 11.61 & 10.61\\\hline
20 & 24.46 & 4.53 & 17.66 & 4.89 & 10.68 & 8.44\\\hline
10 & 25.77 & 4.88 & 18.80 & 6.73 & 11.87 & 8.38\\\hline
5 & 25.02 & 5.45 & 17.02 & 2.98 & 11.65 & 8.78\\\hline
\multicolumn{7}{c}{\bf Sampling: 7 {\AA}}\\\hline
50 & 21.57 & 3.77 & 19.32 & 9.40 & 7.31 & 6.39\\\hline
40 & 23.11 & 6.80 & 17.69 & 8.41 & 7.48 & 6.28\\\hline
30 & 22.97 & 5.89 & 15.43 & 7.15 & 12.83 & 10.04\\\hline
20 & 22.41 & 7.31 & 14.41 & 10.95 & 11.37 & 10.60\\\hline
10 & 23.25 & 4.90 & 19.00 & 1.97 & 6.53 & 6.88\\\hline
5 & 22.98 & 5.48 & 15.82 & 9.18 & 9.44 & 11.51\\\hline
\multicolumn{7}{c}{\bf Sampling: 8 {\AA}}\\\hline
50 & 28.15 & 6.99 & 18.48 & 6.79 & 19.39 & 11.56\\\hline
40 & 26.33 & 6.10 & 17.53 & 7.93 & 18.48 & 13.04\\\hline
30 & 25.27 & 5.51 & 21.04 & 6.04 & 11.13 & 6.68\\\hline
20 & 27.59 & 7.66 & 17.84 & 7.99 & 20.48 & 16.07\\\hline
10 & 27.73 & 6.24 & 16.12 & 9.62 & 23.48 & 15.89\\\hline
5 & 27.23 & 4.66 & 17.84 & 6.26 & 15.99 & 11.97\\\hline
\multicolumn{7}{c}{\bf Sampling: 9 {\AA}}\\\hline
50 & 28.57 & 7.51 & 21.62 & 8.30 & 11.71 & 9.31\\\hline
40 & 27.45 & 8.37 & 21.30 & 9.92 & 13.38 & 10.54\\\hline
30 & 29.69 & 7.24 & 21.94 & 8.57 & 12.47 & 10.39\\\hline
20 & 29.13 & 8.33 & 21.62 & 9.51 & 12.21 & 7.44\\\hline
10 & 27.87 & 9.19 & 20.61 & 9.72 & 12.56 & 9.\\\hline
5 & 28.15 & 7.63 & 23.48 & 6.52 & 9.78 & 7.20\\\hline
\multicolumn{7}{c}{\bf Sampling: 10 {\AA}}\\\hline
50 & 31.10 & 10.52 & 23.09 & 10.55 & 15.22 & 11.24\\\hline
40 & 28.16 & 10.68 & 24.09 & 9.49 & 12.91 & 11.32\\\hline
30 & 29.56 & 9.00 & 24.72 & 10.81 & 13.52 & 8.22\\\hline
20 & 30.12 & 11.90 & 22.46 & 9.60 & 22.95 & 20.73\\\hline
10 & 31.10 & 10.60 & 24.05 & 9.46 & 17.71 & 18.40\\\hline
5 & 30.33 & 9.02 & 25.68 & 9.47 & 15.18 & 11.87\\\hline

\end{tabular} }
\caption{Average H$\alpha$, [{N\,\textsc{ii}}], and [{N\,\textsc{ii}}]/H$\alpha$ error with its respective sigma. Errors were estimated for a tunable filter FWHM of 9{\AA} using the simulated spectra at redshift 0.4}
\end{center}
\end{table}

\end{document}